\documentclass[prd,twocolumn,showpacs,nofootinbib, aps]{revtex4-1}
\usepackage{bm}
\usepackage{graphicx}
\usepackage{multirow}
\usepackage{amsmath}
\usepackage{amssymb}
\usepackage{hyperref}
\usepackage{color}
\usepackage{xspace}
\usepackage{verbatim}
\usepackage{mathtools}
\usepackage{booktabs}
\usepackage[dvipsnames]{xcolor}
\usepackage[caption=false]{subfig}

\newcommand{\vmin}{v_\mathrm{min}}
\newcommand{\qhat}{\hat{\mathbf{q}}}

\newcommand{\kms}{\textrm{ km s}^{-1}}
\newcommand{\dbd}[2]{\ifmmode \frac{\textrm{d}#1}{\textrm{d}#2}\else $\textrm{d}#1/\textrm{d}#2$\fi}

\newcommand{\erf}{\mathrm{erf}}
\newcommand{\vesc}{v_{\rm esc}}

\newcommand{\vy}{\langle v_y \rangle}
\newcommand{\vT}{\sqrt{\langle v_T^2 \rangle}}

\newcommand{\ra}[1]{\renewcommand{\arraystretch}{#1}}

\begin{document}

\title{Reconstructing the three-dimensional local dark matter velocity distribution}

\author{Bradley J. Kavanagh}
\email{bradley.kavanagh@lpthe.jussieu.fr}
\affiliation{LPTHE, CNRS, UMR 7589, 4 Place Jussieu, F-75252, Paris, France}
\affiliation{Institut de physique th\'eorique, Universit\'e Paris Saclay, CNRS, CEA, F-91191 Gif-sur-Yvette, France}

\author{Ciaran A. J. O'Hare}
\email{ciaran.ohare@nottingham.ac.uk}
\affiliation{School of Physics and Astronomy, University of Nottingham, University Park, Nottingham, NG7 2RD, UK}

\begin{abstract}
Directionally sensitive dark matter (DM) direct detection experiments present the only way to observe the full three-dimensional velocity distribution of the Milky Way halo local to Earth. In this work we compare methods for extracting information about the local DM velocity distribution from a set of recoil directions and energies in a range of hypothetical directional and non-directional experiments. We compare a model independent empirical parameterisation of the velocity distribution based on an angular discretisation with a model dependent approach which assumes knowledge of the functional form of the distribution. The methods are tested under three distinct halo models which cover a range of possible phase space structures for the local velocity distribution: a smooth Maxwellian halo, a tidal stream and a debris flow. In each case we use simulated directional data to attempt to reconstruct the shape and parameters describing each model as well as the DM particle properties. We find that the empirical parametrisation is able to make accurate unbiased reconstructions of the DM mass and cross section as well as capture features in the underlying velocity distribution  in certain directions without any assumptions about its true functional form. We also find that by extracting directionally averaged velocity parameters with this method one can discriminate between halo models with different classes of substructure.
\end{abstract}

\maketitle

\section{Introduction}
\label{sec:introduction}
The search for dark matter (DM) via the measurement of keV-scale nuclear recoils in dedicated low-background underground detectors has a unique and potentially powerful directional signature. The relative motion of the Solar system with respect to the non-rotating DM halo of the Milky Way should give rise to an anisotropic flux of DM particles with a peak incoming direction coinciding with the constellation of Cygnus~\cite{Spergel:1987kx}. This peak direction is typically regarded as a `smoking gun' signal for a particle of Galactic origin, as it is not mimicked by any known cosmic or terrestrial background . As such, the measurement of nuclear recoil directions consistent with this predicted direction is a powerful tool for both the discovery of dark matter~\cite{Copi:1999pw,Morgan:2004ys,Billard:2009mf,Green:2010zm,Mayet:2016zxu} as well as continuing the search at cross sections below the neutrino floor~\cite{Grothaus:2014hja,O'Hare:2015mda}. Additionally, directional detection may be the only way of measuring the full three-dimensional velocity distribution of DM at the Earth's Galactic radius~\cite{Billard:2010jh,Billard:2012qu,Lee:2012pf,O'Hare:2014oxa}. This in turn may give insights into the the process of galaxy formation and the merger history of our own Milky Way. For a recent review of the discovery reach of directional detection experiments see Ref.~\cite{Mayet:2016zxu}.

Measuring the direction of nuclear recoils at the keV scale is experimentally challenging. A variety of prototype experiments are currently in operation utilising a range of novel techniques to extract directional information from a nuclear recoil signal (see e.g.,~Refs.~\cite{Cappella:2013,Capparelli:2014lua,Aleksandrov:2016fyr}, as well as Ref.~\cite{Ahlen:2009ev} for a review). One promising approach is to use a gaseous time projection chamber (TPC) at low pressure in order for the track of electrons ionised by a nuclear recoil to be large enough to detect at around $\mathcal{O}(1\,{\rm mm})$ in size. The direction of this recoil can be inferred by drifting the liberated electrons to a time sampled pixelised anode to reconstruct the 3 dimensional orientation of the track. Experiments such as MIMAC~\cite{Riffard:2013psa,Riffard:2016mgw}, DRIFT~\cite{Daw:2011wq,Battat:2014van}, NEWAGE~\cite{Nakamura:2015iza,Miuchi:2010hn}, DMTPC~\cite{Monroe:2011er,Leyton:2016nit} and D3~\cite{Jaegle:2011rn} currently make use of this technology in some variant. Attempts to measure recoil directionality encounter a range of experimental difficulties on top of the usual challenges found in direct detection experiments. The most immediate limitation of gas TPCs is their ability to be scaled to competitive detector masses, with the largest of these prototype experiments currently operating around the 0.1 kg scale~\cite{Santos:2011kf}. There are also challenges that arise in accurately reconstructing the 3-dimensional recoil track. Most notably there is the problem of head-tail recognition - the measurement of the sense of the nuclear recoil (i.e., $+\qhat$ or $-\qhat$) - which has proven to be difficult to achieve~\cite{Billard:2012bk} and has been shown to have a significant impact on the discovery potential of directional experiments~\cite{Green:2007at,Billard:2014ewa}.  

The expected event rate in direct detection (DD) experiments depends crucially on the astrophysics of the local halo. In particular, a failure to properly account for uncertainties in the DM velocity distribution may lead to biased measurements of the DM mass and cross section from a future signal~\cite{Peter:2011eu}. It will therefore be imperative to include these uncertainties in fits to direct detection data. This can be done by fitting to phenomenological models for the local distribution~\cite{Billard:2010jh,Lee:2012pf,O'Hare:2014oxa}, or by attempting to integrate out the astrophysics dependence of the DM signal so that comparisons can be made between exclusion limits from different experiments in a `halo-independent' way~\cite{Fox:2010bz,Fox:2010bu,Frandsen:2011gi,Gondolo:2012rs,DelNobile:2013cta,Fox:2014kua,Feldstein:2014gza,Anderson:2015xaa,Gelmini:2016pei,Kahlhoefer:2016eds}. Alternatively one can use empirical parametrisations of the speed distribution to account for astrophysical uncertainties, although this may lead to weakened constraints on other DM parameters~\cite{Peter:2011eu,Kavanagh:2013wba,Kavanagh:2013eya}.

Here, we extend the use of general parametrisations of the speed distribution to the fitting of the {\it velocity} distribution with directional data.\footnote{We distinguish here between the distribution of the 3-dimensional vector \textit{velocity} $\mathbf{v}$ and the scalar \textit{speed}, given by $v = |\mathbf{v}|$.} Following the formalism introduced in Ref.~\cite{Kavanagh:2015aqa} we will test a binned approach for parametrising the full 3-dimensional local velocity distribution with directional detectors in a model independent way. In this approach the velocity distribution is divided into angular bins, each described by an empirical 1-d speed distribution which does not vary with angle over the bin. The goal of this work is to use mock data to test the accuracy of the reconstucted DM signal using this empirical method compared with model-dependent approaches in both energy only and directionally sensitive direct detection experiments. We compare reconstructions of the DM mass, standard spin-dependent cross section and velocity distribution in three distinct cases: a) when the velocity distribution is known exactly; b) when the general functional form of the distribution is known (as in Refs.~\cite{Billard:2010jh,Lee:2012pf,O'Hare:2014oxa}); and c) when no assumptions are made about the velocity distribution.

To begin in Sec.~\ref{sec:Benchmarks} we will review the relevent directional detection theory and list the benchmark particle and astrophysics parameters that we will attempt to reconstruct. In Sec.~\ref{sec:paramrecon} we describe our mock experimental setups, statistical analysis and methods for reconstructing the velocity distribution. In Sec.~\ref{sec:results} we present the results of reconstructions of the DM mass and cross section as well as the shape and parameters of the velocity distribution. We also include results for directional experiments that lack the ability to tell the forward or backward going sense of observed nuclear recoils. Finally, we discuss the implications of these results and summarise in Sec.~\ref{sec:conc}. 

\section{Benchmarks}
\label{sec:Benchmarks}

\subsection{Particle physics}
The directional DM-nucleus scattering rate per unit detector mass as a function of recoil energy $E_r$ and direction $\qhat$ is given by~\cite{Gondolo:2002np},
\begin{equation}\label{eq:directionalrate}
 \frac{\mathrm{d}^2R}{\mathrm{d}E_r \mathrm{d}\Omega_q} = \frac{\rho_0}{4\pi \mu_{\chi p}^2 m_\chi} \sigma_p \mathcal{C}_N F^2(E_r) \hat{f}(\vmin,\qhat)
\end{equation}
where $m_\chi$ is the DM mass, $\mu_{\chi p}$ is the DM-proton reduced mass and $\sigma_p$ is the DM-proton cross section for either spin-independent (SI) or spin-dependent (SD) interactions. The function $F(E_r)$ is the nuclear form factor parametrising the loss of coherence in the DM-nucleus interaction at high momentum transfer. The coefficient $\mathcal{C}_N$ is an enhancement factor which depends on the nucleon content of the target nucleus $N$, which along with the cross section can encode SI or SD scattering. The velocity distribution enters in the form of its Radon transform $\hat{f}$ at $\vmin = \sqrt{m_N E_r / 2\mu_{\chi N}}$ , the smallest DM speed that can create a recoil of energy $E_r$. Note that the rate given in Eq.~\ref{eq:directionalrate} is valid only for the standard SI/SD contact interactions, which are lowest order in the DM speed $v$. However, it would be possible to extend the analysis to higher-order interactions, such as those of the non-relativistic effective field theory (NREFT) of Fitzpatrick et al.~\cite{Fitzpatrick:2012ix}.

We consider only a single particle physics benchmark in this work, namely a DM particle with a mass of $m_\chi = 50 \, \, \mathrm{GeV}$ and a SD DM-nucleon cross section of $\sigma^{\rm SD}_p = 10^{-39} \, \, \mathrm{cm}^2$. We assume that the DM particle has no SI coupling and that the ratio of couplings to protons and neutrons is $a_p/a_n = -1$ \cite{Jungman:1995df}. This benchmark is not currently excluded by constraints\footnote{Note that direct detection constraints typically assume couplings only to protons or neutrons. In our case, we couple to both (with opposite signs). This typically leads to a slight cancellation in the event rate and therefore weaker constraints on our benchmark model compared to those reported by experimental collaborations.} from either direct detection~\cite{Amole:2015pla,Amole:2015lsj,Akerib:2016lao} or neutrino telescope searches~\cite{Aartsen:2016exj} (assuming annihilation into soft channels such as $b \overline{b}$). Furthermore this benchmark gives a sizeable rate in both Xenon and Fluorine targets, allowing us to explore the complementarity of multiple directionally-sensitive experiments. This is because Xenon has a reduced sensitivity to DM-proton SD interactions, having most of its spin carried by an unpaired neutron. By comparison we use a Fluorine experiment with a smaller exposure that is compensated by its greater sensitivity to DM-proton SD interactions.

\subsection{Astrophysics}
The scattering rate is dependent on the Earth frame DM velocity distribution $f(\mathbf{v})$ in the form of its Radon transform~\cite{Gondolo:2002np},
\begin{equation}
 \hat{f}(\vmin,\qhat) = \int f(\mathbf{v}) \delta(\mathbf{v}\cdot\qhat-\vmin)\, \mathrm{d}^3\mathbf{v} \, .
\end{equation}
Most direct detection analyses are performed under a simple assumption for the Milky Way halo known as the standard halo model (SHM)~\cite{Green:2011bv}. This is a spherically symmetric isothermal halo model with a $1/r^2$ density profile, yielding a Maxwell-Boltzmann (MB) velocity distribution. The SHM is now a commonplace assumption and a number of recent hydrodynamic simulations suggest that a simple MB distribution is sufficient to describe the local velocity distribution \cite{Bozorgnia:2016ogo,Kelso:2016qqj,Sloane:2016kyi}. However, other hydrodynamic simulations (as well as earlier N-body simulations) present evidence that the SHM may not accurately reflect the true Milky Way halo~\cite{Lisanti:2010qx,Kuhlen:2012ft,Fornasa:2013iaa,Butsky:2015pya}. The matter has not yet been conclusively settled and, critically for direct detection experiments, this means that the local velocity distribution at the Earth's Galactic radius may contain significant departures from a Maxwellian form~\cite{Vogelsberger:2008qb,Maciejewski:2010gz,Mao:2012hf}. The distribution may also contain additional features and substructures such as debris flows~\cite{Lisanti:2011as,Kuhlen:2012fz}, tidal streams~\cite{Freese:2003tt,Purcell:2012sh}, a co-rotating dark disk~\cite{Read:2009iv,Kuhlen:2013tra,Schaller:2016uot} or a `Shadow Bar'~\cite{Petersen:2016xtd,Petersen:2016vck}.

We consider three astrophysical benchmarks in this work which are motivated by results from N-body simulations, but also importantly have very different velocity structures so that the different approaches for reconstructing the velocity distribution can be compared under a range of scenarios. These distributions are: \\

\textbf{Standard Halo Model (SHM)}: The SHM has a MB distribution, with peak speed $v_0 = 220 \kms$ and width $\sigma_v = v_0/\sqrt{2} \approx 156 \kms$. The Earth's speed is set equal to the peak speed and we fix the escape speed to the best fit RAVE measurement $\vesc = 533 \kms$~\cite{Piffl:2013mla}. The velocity distribution in the Earth frame is therefore given by:
\begin{eqnarray}
f_\mathrm{SHM}(\mathbf{v}) =& \frac{1}{(2\pi \sigma_v^2)^{3/2}N_\mathrm{esc}} \, \exp \left( - \frac{(\mathbf{v} - \mathbf{v}_0)^2}{2\sigma_v^2}\right) \\
&\times \, \Theta (\vesc - |\mathbf{v} - \mathbf{v}_0|)\,, \nonumber
\end{eqnarray}
with the normalisation constant given by
\begin{equation}
N_\mathrm{esc} = \erf \left( \frac{\vesc}{\sqrt{2}\sigma_v}\right) - \sqrt{\frac{2}{\pi}} \frac{\vesc}{\sigma_v} \exp \left( -\frac{\vesc^2}{2\sigma_v^2}   \right)\,.
\end{equation}
To define velocities we use the Galactic co-ordinate system in which the Earth's velocity points in the $y$-direction i.e., $\textbf{v}_0 = (0,v_0,0)$.\\

\textbf{SHM + Stream (SHM+Str)}: The local velocity distribution may also contain substructure from the tidal disruption of satellite galaxies of the Milky Way. There is some evidence that the tidal stripping of material from the nearby Sagittarius dwarf galaxy could pass through the Earth's location~\cite{Purcell:2012sh}. Due to the spatially and kinematically localised nature of these features they give rise to prominent directional signatures in the recoil spectrum~\cite{Lee:2012pf,O'Hare:2014oxa}. We assume that a fixed fraction of the local density is contained in the form of a tidal stream, described by Galactic frame velocity $\mathbf{v}_\mathbf{s}$ and dispersion $\sigma_\mathrm{s}$. The velocity distribution of the stream is given by,
\begin{equation}
f_\mathrm{Str}(\mathbf{v}) = \frac{1}{(2\pi \sigma_\mathrm{s}^2)^{3/2}} \, \exp \left( - \frac{(\mathbf{v} - (\mathbf{v}_0-\mathbf{v}_\mathrm{s}))^2}{2\sigma_s^2}\right) \,,
\end{equation}
and the full velocity distribution of the ``SHM+Str'' model is given by,
\begin{equation}
 f_\mathrm{SHM+Str}(\mathbf{v}) = \left(1-\frac{\rho_\mathrm{s}}{\rho_0}\right)f_\mathrm{SHM}(\mathbf{v}) + \frac{\rho_\mathrm{s}}{\rho_0} f_\mathrm{Str}(\mathbf{v}) \, .
\end{equation}
where $\rho_0$ is the SHM density and $\rho_\mathrm{s}$ is the stream density. \\

\textbf{SHM + Debris Flow (SHM+DF)}: Debris flows are another form of substructure that has been seen to appear in N-body simulations such as Via Lactea II~\cite{Kuhlen:2008qj,Kuhlen:2012fz}. Like streams these are kinematically localised, characterised by a speed $v_f$, though unlike streams they are spatially extended features which form from the incomplete phase mixing of material during the formation of the halo. Following Ref.~\cite{Kuhlen:2012fz} we assume a model for the debris flow in which the velocity distribution is isotropic in the Galactic frame and a delta function in speed centered on $v_f$,
\begin{equation}
 f_\mathrm{DF}(\mathbf{v}) = \frac{1}{4\pi v_f^2} \,\delta(|\mathbf{v}-\mathbf{v}_0|-v_f) \, .
\end{equation}
As with the SHM+Str model we combine the debris flow with the SHM as a fixed fraction of the local density:
\begin{equation}
 f_\mathrm{SHM+DF}(\mathbf{v}) = \left(1-\frac{\rho_f}{\rho_0}\right)f_\mathrm{SHM}(\mathbf{v}) + \frac{\rho_f}{\rho_0} f_\mathrm{DF}(\mathbf{v}) \, .
\end{equation}

\begin{figure}[t]
\includegraphics[width=0.48\textwidth,trim={0 4.2cm 0 0}]{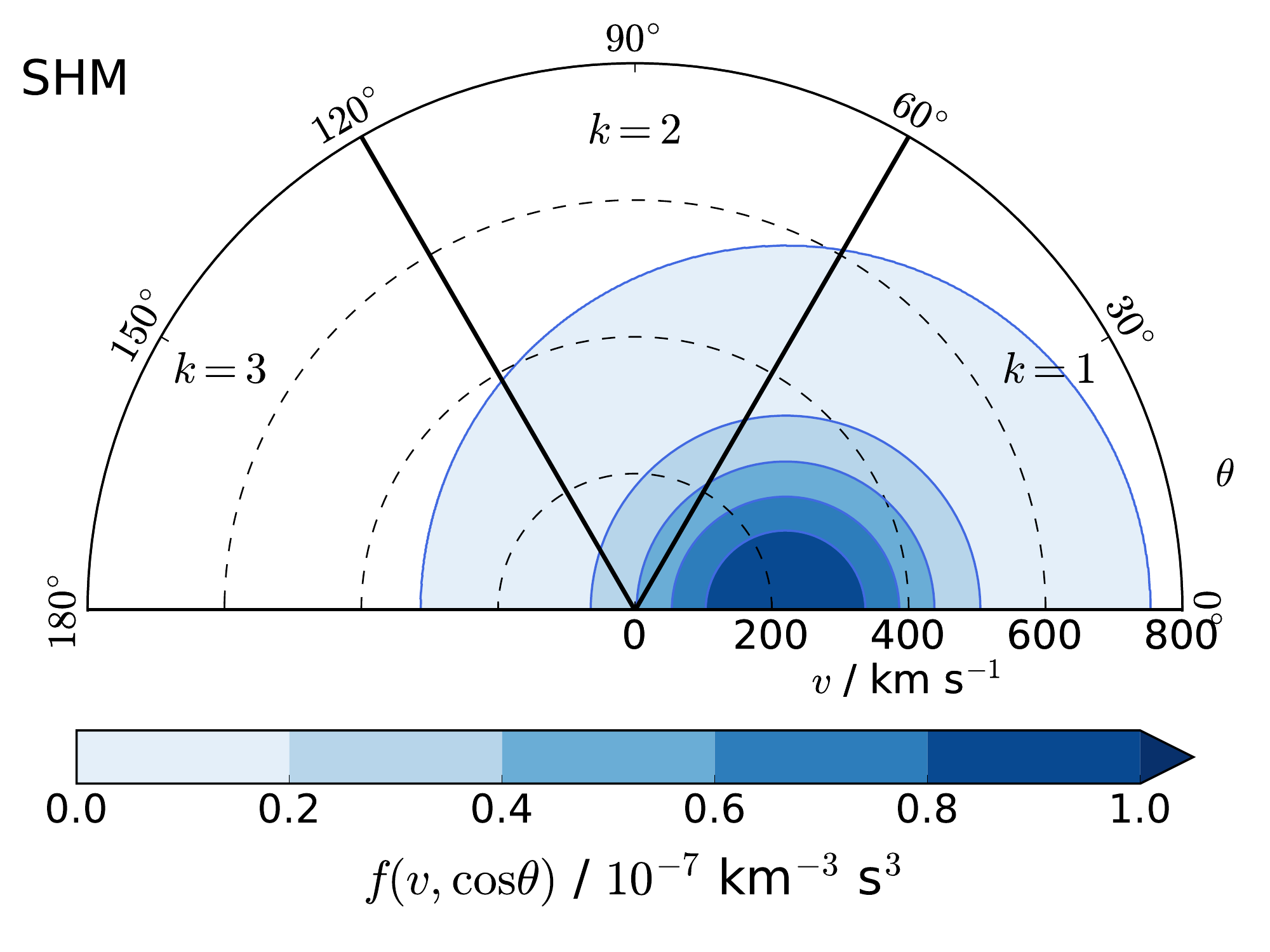}
\includegraphics[width=0.48\textwidth,trim={0 4.2cm 0 0}]{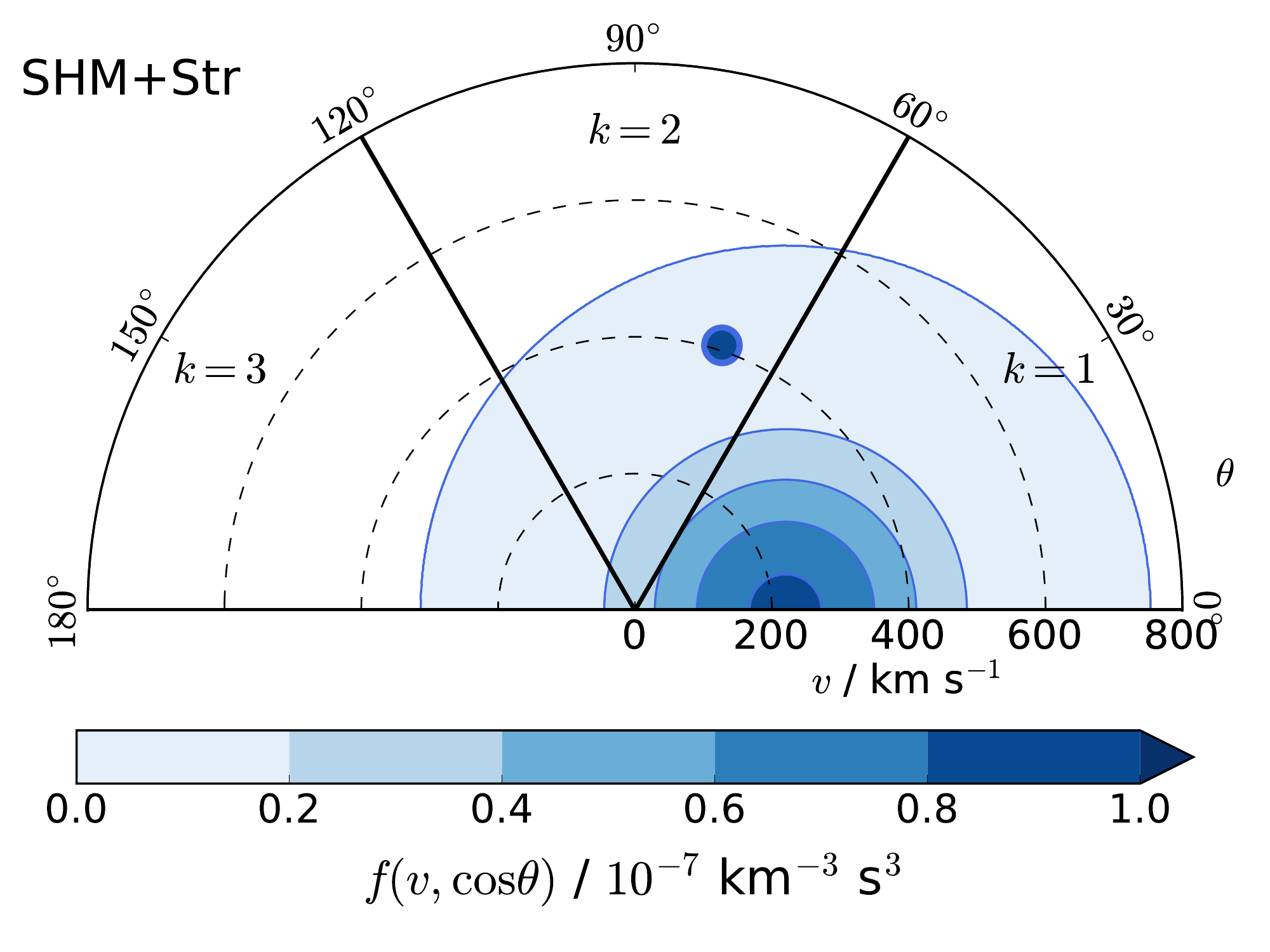}
\includegraphics[width=0.48\textwidth]{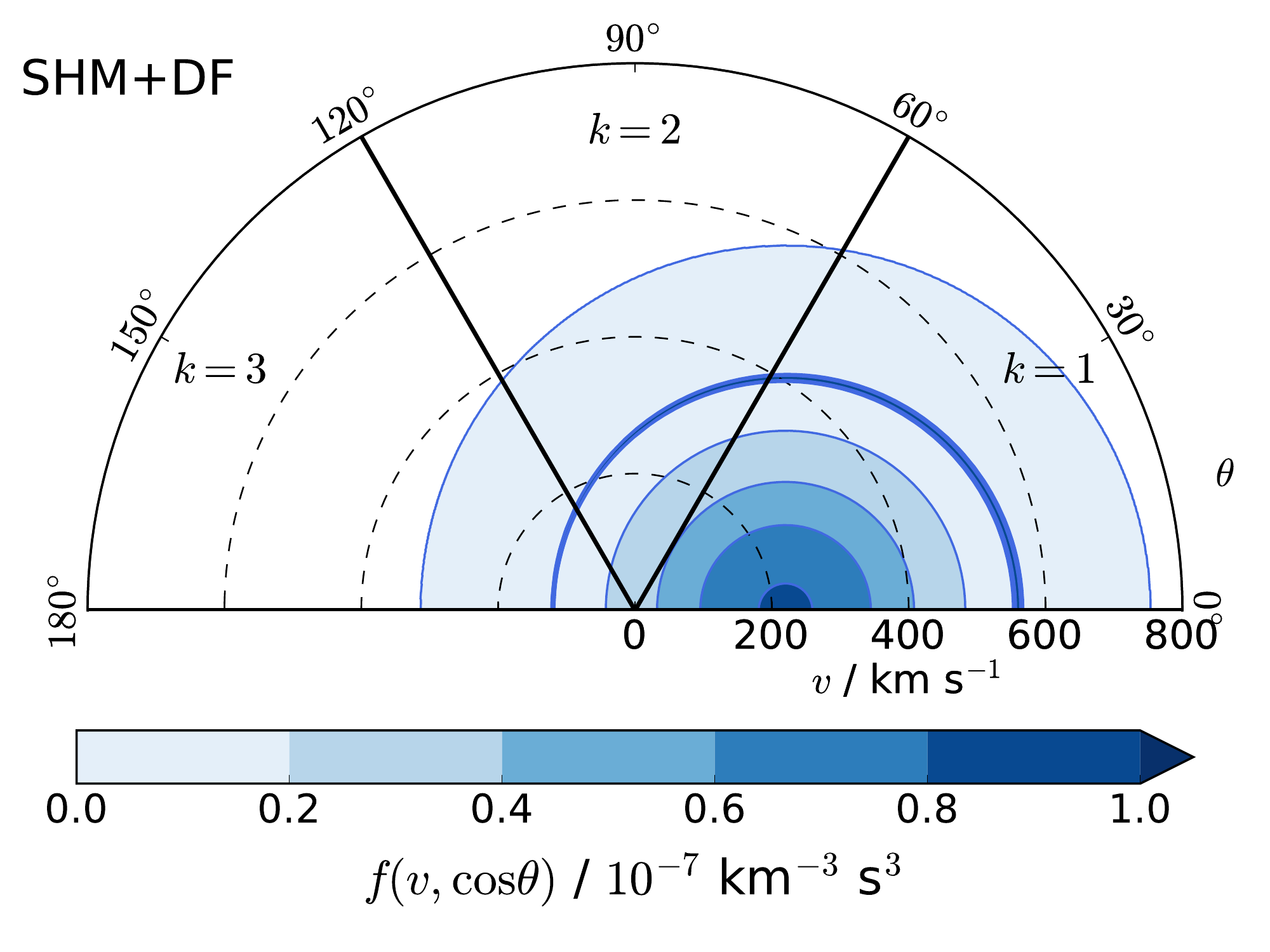}
\caption{Benchmark velocity distributions used in the analysis. We plot the velocity distribution for the SHM (top), SHM + Stream (middle) and SHM + Debris Flow (bottom). The polar angle $\theta$ is measured with respect to $\mathbf{v}_0$ and we have integrated over the azimuthal angle $\phi$. We also label the angular bins ($k=1,2,3$) used in the empirical parametrisation (see Sec.~\ref{sec:discretisation}).}\label{fig:polar}
\end{figure}

These benchmark velocity distributions are shown in Fig.~\ref{fig:polar}, while a summary of the benchmark parameter values used for each halo model is given in Table~\ref{tab:astrobenchmarks}. For the stream we use an estimate of the velocity of the Sagittarius stream from Ref.~\cite{Savage:2006qr}. However we assume that it comprises a significantly larger fraction of the local density than suggested by simulations, typically around the 1\% level~\cite{Vogelsberger:2008qb,Maciejewski:2010gz}. This allows us to make a clear distinction between our benchmark models. For the debris flow we use the parameters derived in the semi-analytic model of Ref.~\cite{Kuhlen:2012fz} based on the Via Lactea II simulation~\cite{Kuhlen:2008qj}. Although the debris flow in this simulation exhibited some velocity dispersion as well as a small bias towards directions tangential to the Galactic rotation, the simple isotropic model was found to capture the main features of the recoil spectrum.

\begin{table}[t!]
\ra{1.3}
\begin{tabular}{@{}lll@{}}
\hline\hline
\multirow{4}{*}{\bf SHM} & $\rho_0$ 		& $0.3 \, \mathrm{GeV} \, \mathrm{cm}^{-3}$\\
				& $v_0$ 	& $220 \kms$\\
				& $\sigma_v$ 		& $156 \kms$\\
				& $\vesc$ & $533 \kms$\\
\hline
\multirow{3}{*}{\quad{\bf +Str}} & $\sigma_s$ 		& $10 \, \kms$\\
				& $\textbf{v}_s$ 	& $400\times(0,0.233,-0.970) \kms$\\
				& $\rho_s/\rho_0$ 		& $0.2$\\
\hline
\multirow{2}{*}{\quad{\bf +DF}} & $v_f$ 		& $340 \, \kms$\\
				& $\rho_f/\rho_0$ 		& $0.22$\\
\hline\hline
\end{tabular}
\caption{Astrophysical benchmark parameters for the three halo models under consideration: the standard halo model alone, and with the addition of a stream and debris flow.}
\label{tab:astrobenchmarks}
\end{table}

In all cases, we neglect any time dependence of the Earth's velocity (which may lead to a percent-level modulation of the event rate~\cite{Drukier:1986tm}) and we fix the local DM density to $\rho_0 = 0.3 \, \, \mathrm{GeV} \, \, \mathrm{cm}^{-3}$. Local and global estimates of the local DM density give an uncertainty of roughly a factor of 2 (for a review, see Ref.~\cite{Read:2014qva}). However, the DM density is common to all experiments and this uncertainty is degenerate with the DM-nucleon cross section.

\section{Parameter Reconstruction}
\label{sec:paramrecon}

\subsection{Mock experiments}
\label{sec:experiments}
We consider two ideal background-free mock experiments: a Xenon-based experiment and a Fluorine-based experiment. The SD nuclear enhancement factor (appearing in Eq.~(\ref{eq:directionalrate})) for each target can be written in terms of the expectation values of the proton and neutron spin $\langle S_{p,n}\rangle$ and total nuclear spin $J$~\cite{Engel:1992bf},
\begin{equation}
 \mathcal{C}^{\mathrm{SD}}_N = \frac{4}{3} \frac{J+1}{J} | \langle S_p \rangle + a_n/a_p \langle S_n \rangle |^2 .
\end{equation}
The nuclear spin parameters used for Xe and F are shown in Table~\ref{tab:nuclear}. For Xenon, we use the Menendez {\it et al}.~SD structure functions \cite{Menendez:2012tm}, absorbing the two-body corrections into the values of the spin content. For Fluorine, we use the Divari {\it et al}.~structure functions \cite{Divari2000}, including the two-body corrections as reported by Cannoni \cite{Cannoni:2012jq}. 


\begin{table}[t!]\centering
\begin{tabular}{@{}lcccc@{}}
\hline \hline
				& J		& $\langle S_p \rangle$	& $\langle S_n \rangle$ & Isotopic fraction \\
				\hline
$^{19}\mathrm{F}$ 	& 1/2		& 0.421				& 0.045 & 1 \\
$^{129}\mathrm{Xe}$ & 1/2	& 0.046 				& 0.293 & 0.265\\
$^{131}\mathrm{Xe}$ & 3/2	& -0.038				& -0.242 & 0.212\\
\hline \hline
\end{tabular}
\caption{Spin content and abundances of the targets considered in the two mock experiments.}
\label{tab:nuclear}
\end{table}

In addition to considering two different target nuclei, we vary the amount of directional information used in the reconstruction, considering cases in which neither experiment has directional sensitivity, in which only one of the experiments has directional sensitivity and in which both experiments are directionally sensitive. A summary of the parameters used for each experiment and the number of events observed in those experiments for each halo model are given in Table~\ref{tab:benchmarks}. The choice of a Xenon target detector is inspired by projections for the next generation of ton-scale liquid Xenon experiment such as LZ~\cite{Akerib:2015cja} and Xenon1T~\cite{Aprile:2015uzo}. Although these experiments are not designed with any directional sensitivity they represent a useful and realistic benchmark for an exposure and threshold ($\sim 5\,\textrm{keV}$) that can be expected in the next generation of direct detection experiments. However we note that there are tentative suggestions that it may be possible to extract directional information in liquid Xenon experiments with existing technology by exploiting columnar recombination~\cite{Nygren:2013nda,MuAaoz:2014uxa,Li:2015zga,Mohlabeng:2015efa}. The choice of a Fluorine detector is inspired by existing low pressure gas TPCs with CF$_4$ such as NEWAGE~\cite{Nakamura:2015iza,Miuchi:2010hn} and DMTPC~\cite{Leyton:2016nit} although we extend to a 10 kg mass so that the sensitivity of the experiment reaches beyond currently excluded regions of the spin-dependent DM parameter space. We set a typical threshold of 20 keV, in line with what is currently achievable~\cite{Leyton:2016nit}.

\begin{table}[t!]
\ra{1.3}
\begin{tabular}{@{}lll@{}}
\hline \hline
				& Expt 1		& Expt 2 \\
				\hline 
Target			& Xe 			& F \\
$E_\mathrm{th}$/keV 	& 5			& 20 \\
$E_\mathrm{max}$/keV  & 50			& 50 \\
$\mathcal{E}$/kg yr 		& 1000		& 10 \\
\hline
$N_\mathrm{events}^{\mathrm{SHM}}$&	878		&	50	\\
$N_\mathrm{events}^{\mathrm{SHM+Str}}$&	922		&	67	\\
$N_\mathrm{events}^{\mathrm{SHM+DF}}$&	893		&	64	\\
\hline \hline
\end{tabular}
\caption{Parameters for the two mock experiments considered in this work: threshold energy $E_\mathrm{th}$, maximum analysis energy $E_\mathrm{max}$ and exposure $\mathcal{E}$. Also shown are the number of expected events in the two experiments for each of the three astrophysical benchmarks.}
\label{tab:benchmarks}
\end{table}

\subsection{Statistical analysis}
\label{sec:analysis}
We use a maximum likelihood parameter estimation method to reconstruct the input DM mass, cross section and relevant velocity distribution parameters. We calculate the background-free unbinned extended likelihood, which is the product of the probability of observing each event at its energy and direction, multiplied by the Poisson probability of obtaining the observed number of events in each experiment. When we assume that a given experiment has no directional sensitivity, we simply discard the directional information about each event and use only the recoil energy in the fit.

We use only three sets of mock data (that is, one dataset for each halo model). Whilst any single Poisson realisation of the data will lead to slightly inaccurately reconstructed parameters, given that the total number of events for each halo model is relatively high ($\sim 1000$) these errors are small. Additionally since we are concerned with comparing methods of reconstructing parameters using the same dataset, having multiple Poisson realisations will not affect the conclusions. To explore the parameter space we use the nested sampling algorithms provided by the {\sc MultiNest} package~\cite{Feroz:2007kg,Feroz:2008xx}. In each case we use 10000 live points and a tolerance of $10^{-3}$. 

The DM particle and velocity distribution reconstructions are attempted with three methods each with a different level of a priori knowledge assumed.
\begin{itemize}
\item{{\bf Method A: Perfect knowledge.} This is the best case scenario when both the functional form and parameter values of the velocity distribution are known exactly. The parameters that are reconstructed with this method are only $\{m_\chi,\sigma_p^{\textrm{SD}}\}$ for all three halo models. We place log-flat priors on both parameters in the range $[0.1,1000]$~GeV for $m_\chi$ and $[10^{-40},10^{-37}]\,\textrm{cm}^2$ for $\sigma_p^{\textrm{SD}}$.}

\item{{\bf Method B: Functional form known.} In this case the functional form of the velocity distribution (i.e., SHM, SHM+Str or SHM+DF) is known, however the parameter values are not. The number of parameters reconstructed with this mathod varies depending on the chosen halo model. In the case of the SHM there are 4 parameters: $\{m_\chi,\sigma_p^{\textrm{SD}},v_0,\sigma_v\}$. For the SHM+Str model there are 9 parameters: $\{m_\chi,\sigma_p^{\textrm{SD}},v_0,\sigma_v,\sigma_s,\mathbf{v}_s,\rho_s\}$, and for the SHM+DF model there are 6 parameters: $\{m_\chi,\sigma_p^{\textrm{SD}},v_0,\sigma_v,v_f,\rho_f\}$. We neglect the parameter $\vesc$ which has a negligible effect on the velocity distribution at the energies we are studying and is very difficult to constrain with direct detection data. For each velocity parameter we sample from flat priors in the range $[0,500] \kms$ and for the density of the stream and debris flow we set flat priors in the range $[0,\rho_0]$.}

\item{{\bf Method C: Empirical parametrisation.} With this method no knowledge is assumed about the form or parameters of the underlying velocity distribution. We fit the data using a discretised velocity distribution with $N=3$ angular bins. This method is described in more detail in Sec.~\ref{sec:discretisation}. Three parameters are used to describe the speed distribution within each angular bin, for a total of 11 parameters: $\{ m_\chi, \sigma_p^{\textrm{SD}}, a_0^{(k=1)}, a_1^{(k=1)}, \ldots, a_2^{(k=3)}, a_3^{(k=3)} \}$.} Each of the $a_m^{(k)}$ parameters is sampled linearly in the prior range $[-20, 20]$.

\end{itemize}

\subsection{Discretised velocity distribution}
\label{sec:discretisation}

To perform the model-independent reconstruction (Method C), we discretise the velocity distribution into $N$ angular bins, assuming that $f(\mathbf{v})$ has no angular dependence within each bin. As discussed in Ref.~\cite{Kavanagh:2015aqa}, using only $N=2$ angular components does not sufficiently capture the directionality of typical velocity distributions. We therefore use $N=3$ angular bins, such that the approximate velocity distribution \textit{in the Earth frame} can be written:
\begin{equation}\label{eq:discretisedf}
f(\textbf{v}) = f(v, \cos\theta, \phi) =
\begin{cases}
f^1(v) & \textrm{ for } \theta \in \left[ 0, \frac{\pi}{3}\right]\,, \\
f^2(v) & \textrm{ for } \theta \in \left[ \frac{\pi}{3}, \frac{2\pi}{3}\right]\,, \\
f^3(v) & \textrm{ for } \theta \in \left[ \frac{2\pi}{3}, \pi\right]\,. \\
\end{cases}
\end{equation}
We align the angular bins such that $\theta = 0$ (the `forward' direction) points along $\mathbf{v}_0$, anticipating that the greatest anisotropy in the velocity distribution will be generated by the motion of the Earth through the halo. In Fig.~\ref{fig:polar} we display the three benchmark velocity distributions used in this study, where we also label the bins $k=1,2,3$ used for the discretisation.



The advantage of a discretised velocity distribution is that provided a suitable parameterisation for each $f^k(v)$ is chosen then the complete $f(\textbf{v})$ can be ensured to be everywhere positive, properly normalised and does not require any assumptions about the equilibrium conditions of the Milky Way halo. These issues are often not addressed by other attempts to describe $f(\mathbf{v})$ such as those using functions of integrals of motion~\cite{Alves:2012ay} or decompositions into spherical harmonics and Fourier-Bessel functions~\cite{Lee:2014cpa}.

Within each bin, we follow Ref.~\cite{Kavanagh:2013wba} and describe the 1-d (directionally averaged) velocity distributions using the following empirical parametrisation,
\begin{equation}\label{eq:polynomialparam}
f^k(v) = \exp \left[ - \sum_{m = 0}^3 a_m^{(k)} P_m(2v/v_\mathrm{max} - 1)\right]\,.
\end{equation}
Here, $P_m$ is the $m$th Chebyshev polynomial of the first kind. A value of $v_\mathrm{max} = 1000 \kms$ is chosen as a conservative cut-off for the velocity distribution. The shape of the velocity distribution within each bin is controlled by the parameters $\{a_m^{(k)}\}$. The values of $a_0^{(k)}$ are fixed by requiring that $f^k(0)$ is the same for all $k$ (i.e.~that the three distributions are consistent as we move towards the value $\mathbf{v} = 0$). Finally, we rescale each of the $a_0^{(k)}$ in order to ensure that the full distribution is normalised to unity. This leaves us with three parameters in each of the $N=3$ angular bins, for a total of 9 parameters describing the velocity distribution.

The calculation of the Radon Transform from this discretised distribution is detailed in Ref.~\cite{Kavanagh:2015aqa}. When fitting the parameters of this empirical distribution, we do not keep all of the directional information for each event but instead bin the data into three angular bins (the same angular bins as defined in Eq.~\ref{eq:discretisedf}, but with $\theta$ now referring to the \textit{nuclear recoil angle} with respect to $\mathbf{v}_0$). Within each angular bin in the data, we calculate the extended likelihood using only the recoil energies of the events.  The expected recoil spectrum (as a function of $E_R$) is calculated by integrating the Radon Transform $\hat{f}(v_\mathrm{min}, \hat{\mathbf{q}})$ over the relevant angular range. For example, in the $j$th angular recoil bin, the differential rate of recoils (as a function of energy) is proportional to:
\begin{equation}
\label{eq:discreteRadon}
\hat{f}^j(v_\textrm{min}) = \int_{\phi = 0}^{2\pi} \int_{\cos(j\pi/N)}^{\cos((j-1)\pi/N)} \hat{f}(v_\textrm{min}, \hat{\textbf{q}})\, \mathrm{d}\cos\theta\mathrm{d}\phi\,,
\end{equation}
where $\theta$ and $\phi$ now refer to the direction of the recoil.

There are two reasons for this binning of the data. First, the full Radon Transform of this coarsely discretised distribution is unlikely to give a good fit to the distribution of recoil directions on an event-by-event basis. Instead, if we bin the data on a similar angular scale (or equivalently, integrate the rate over angular bins), this should eliminate any spurious features in the directional spectrum and help mitigate the error induced by using such a discretised approximation. Second, integrating the rate over angular bins allows the angular integrals in the calculation of the Radon Transform to be performed analytically.

\section{Results}
\label{sec:results}

We now present the reconstructed intervals for the particle physics parameters $m_\chi$ and $\sigma_p^\mathrm{SD}$, for the shape of the velocity distribution, and for a number of derived parameters which characterise the anisotropy and width of the velocity distribution. For each reconstruction, the best-fit point is given by the parameter values which maximise the likelihood. We then construct (1- or 2-dimensional) confidence intervals around this point by calculating the profile likelihood and using the asymptotic properties of the profile likelihood ratio \cite{Cowan:2010js}.

\subsection{DM mass and cross section}
\label{sec:mass}

\begin{figure*}[t]
\includegraphics[width=0.48\textwidth]{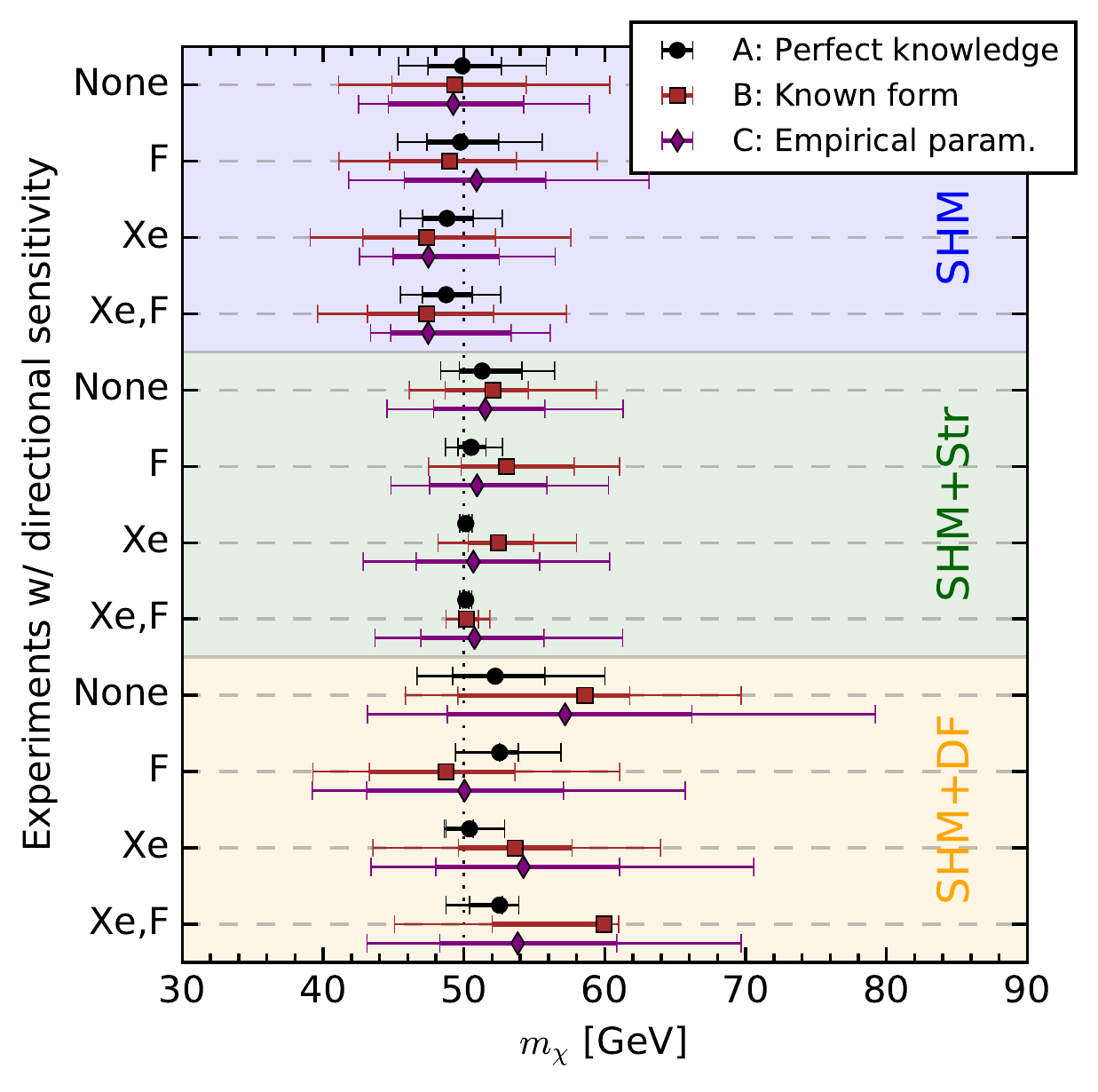}
\includegraphics[width=0.47\textwidth]{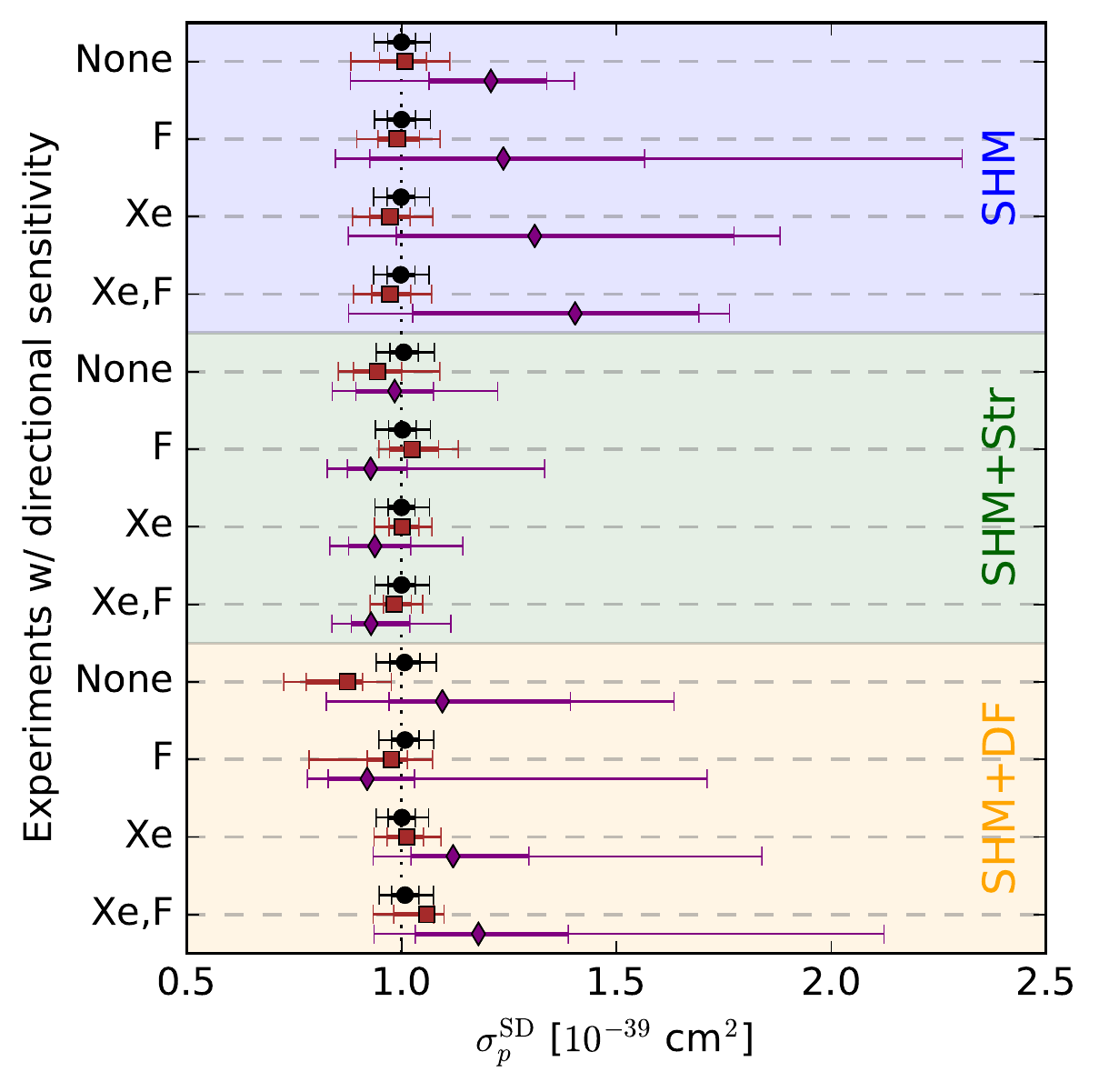}
\caption{Reconstructed 68\% and 95\% confidence intervals for DM mass (left panel) and DM-nucleon SD cross section (right panel) under each halo model (from top to bottom): the SHM (blue region), the SHM with stream (green) and SHM with debris flow (yellow) models. The intervals are shown as a function of the amount of directional information included. The black points and error bars show the reconstruction using perfect knowledge of the DM distribution (Method A), dark red squares show reconstructions when the functional form is known (Method B), and purple diamonds when a general empirical form for the speed distribution is assumed. The input values of the DM mass (50 GeV) and SD cross section ($10^{-39}$ cm$^2$) are shown as vertical dotted lines.}\label{fig:mx-recon}
\end{figure*}

To begin, in the left panel of Fig.~\ref{fig:mx-recon}, we compare the reconstruction of the DM mass using each of the three approaches. In the best case scenario (Method A) when the velocity distribution is known exactly, the WIMP mass is reconstructed with high accuracy, obtaining best fit values with less than 2\% deviation from the input value of $m_\chi = 50 \,\,\mathrm{GeV}$. 

Generally with less assumed knowledge the error on the reconstructed DM mass is larger. However in the case of the SHM the constraints are wider in Method B than in Method C. This is likely due to the small (4-dimensional) parameter space used to reconstruct the SHM. The greater freedom in the (11-dimensional) empirical parametrisation (Method C) may allow for a better fit to the data in the presence of Poisson noise, leading to tighter constraints. For the SHM+Str and SHM+DF models, the underlying velocity distributions are more complex and the parameter space is much larger (9 and 6 dimensions respectively). In these models, the known functional form of Method B can fit the data closely. The empirical parametrisation instead  explores a wide range of the parameter space, but cannot resolve the fine-grained features of these models, leading to wider uncertainties.

We note that using each of the three methods, the true value of the DM mass lies within the 68\% confidence interval in all cases. The best fit DM masses reconstructed using Methods B and C are typically close in value, indicating that there is little bias induced in using the empirical parametrisation, despite the fact that we have used a discretised approximation to $f(\mathbf{v})$ and have assumed very little about the shape of the underlying distribution.

In the right panel of Fig.~\ref{fig:mx-recon}, we show the corresponding limits on the DM-proton SD cross section. In this case, the contrast between Methods A \& B and Method C is more stark. Using the former two methods, reconstruction of $\sigma_p^\mathrm{SD}$ is relatively precise, with an uncertainty of less than 10\%. However, for Method C, the intervals are much wider, extending in most cases up to large values of the cross section. This results from a known degeneracy between the DM cross section and the shape of the speed distribution \cite{Kavanagh:2013wba} in halo-independent approaches. An increase in the fraction of low-speed particles below the direct detection threshold has no effect on the event rate, provided the value of the cross section is increased to counteract the reduced fraction of high-speed particles.\footnote{Note that this degeneracy could be broken if a signal of DM annihilation in the Sun were observed using a neutrino telescope. Low-speed DM particles are captured preferentially by the Sun, leading to complementarity with direct detection \cite{Kavanagh:2014rya,Blennow:2015oea,Ferrer:2015bta}.}

For Method A and B we see that in most cases increasing the quantity of directional information (reading Fig.~\ref{fig:mx-recon} from top to bottom in each halo model) leads to better measurements of the DM mass. In contrast, the error on $\sigma_p^{\rm SD}$ found with Methods A and B is largely insensitive to the amount of directionality as the key information for reconstructing a cross section is the total number of events. For Method C, there is little increase in precision as the amount of directional information is increased; reconstruction of the DM mass in this case depends primarily on obtaining the correct distribution of recoil energies in each experiment. 

\subsection{Velocity distribution shape}
\label{sec:shape}

\begin{figure*}
	\subfloat[\textbf{SHM benchmark. Directionality in F only.}]{\label{fig:veldist-a}
	\includegraphics[width=0.48\textwidth]{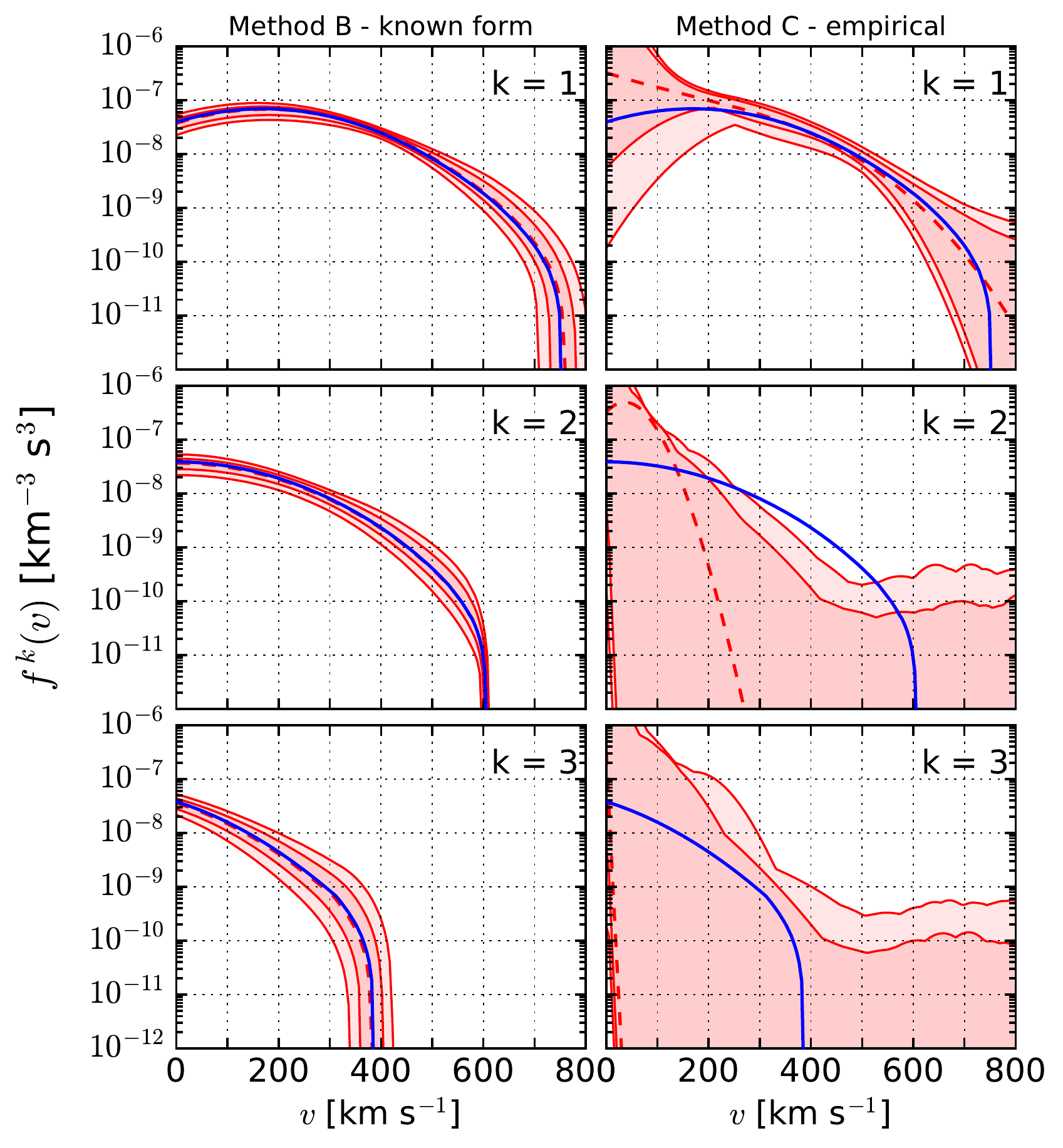}}
	\subfloat[\textbf{SHM benchmark. Directionality in F and Xe.}]{ \label{fig:veldist-b}
		\includegraphics[width=0.48\textwidth]{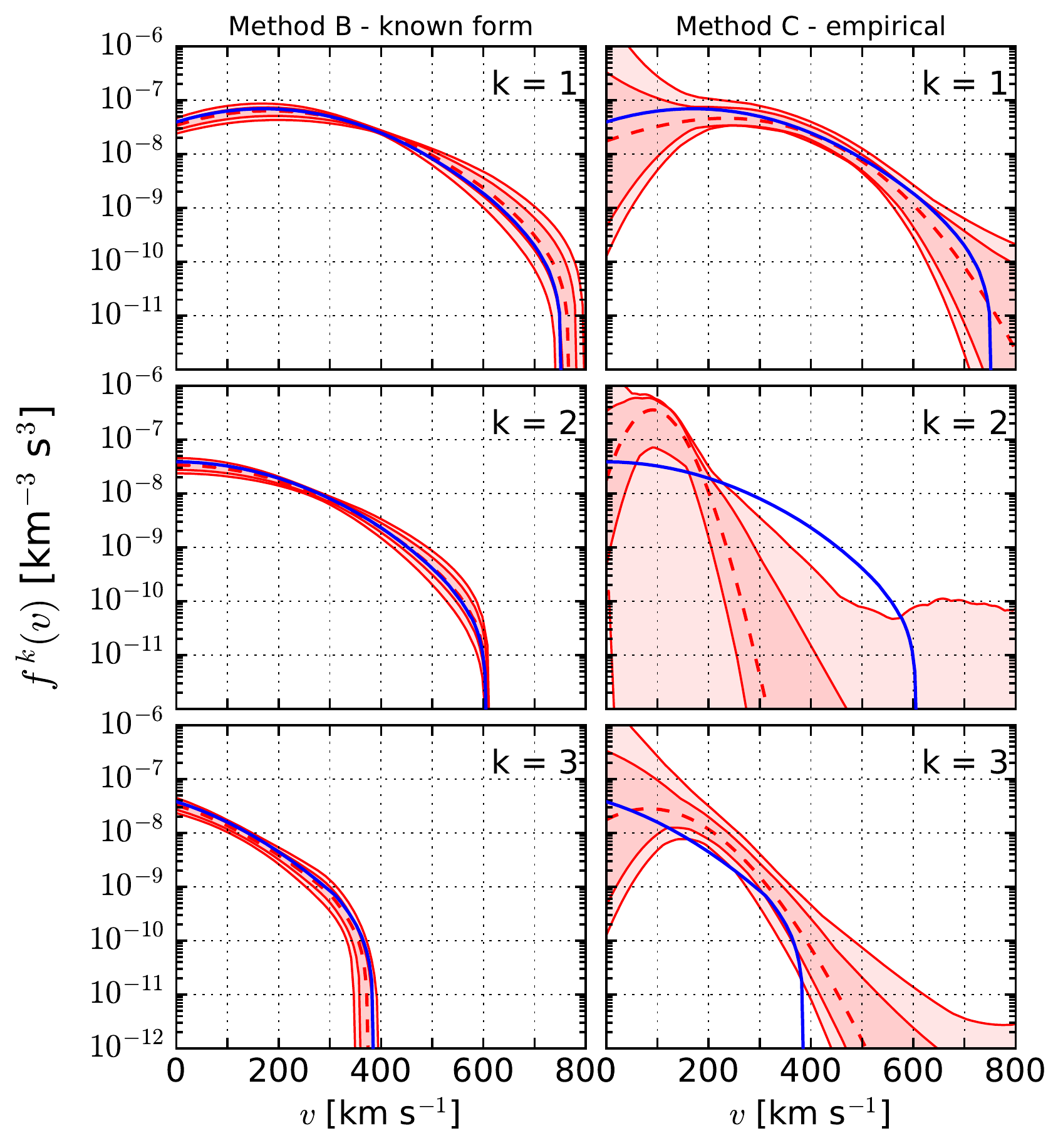}}
	
	\vspace{0.5cm}
	
	\subfloat[\textbf{SHM+Str benchmark. Directionality in F and Xe.}]{ \label{fig:veldist-c}
		\includegraphics[width=0.48\textwidth]{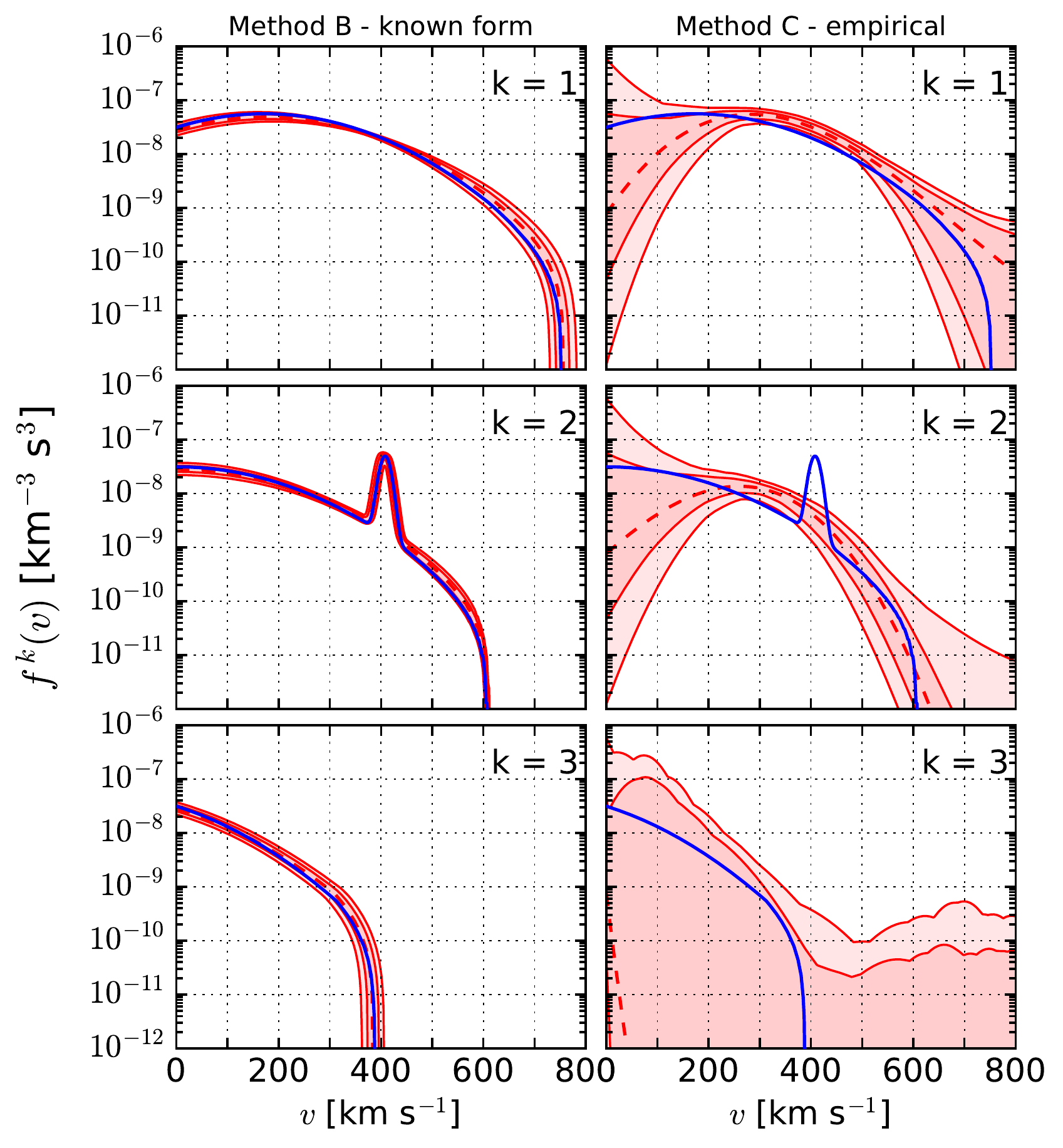}}
	\subfloat[\textbf{SHM+DF benchmark. Directionality in F and Xe.}]{ \label{fig:veldist-d}
		\includegraphics[width=0.48\textwidth]{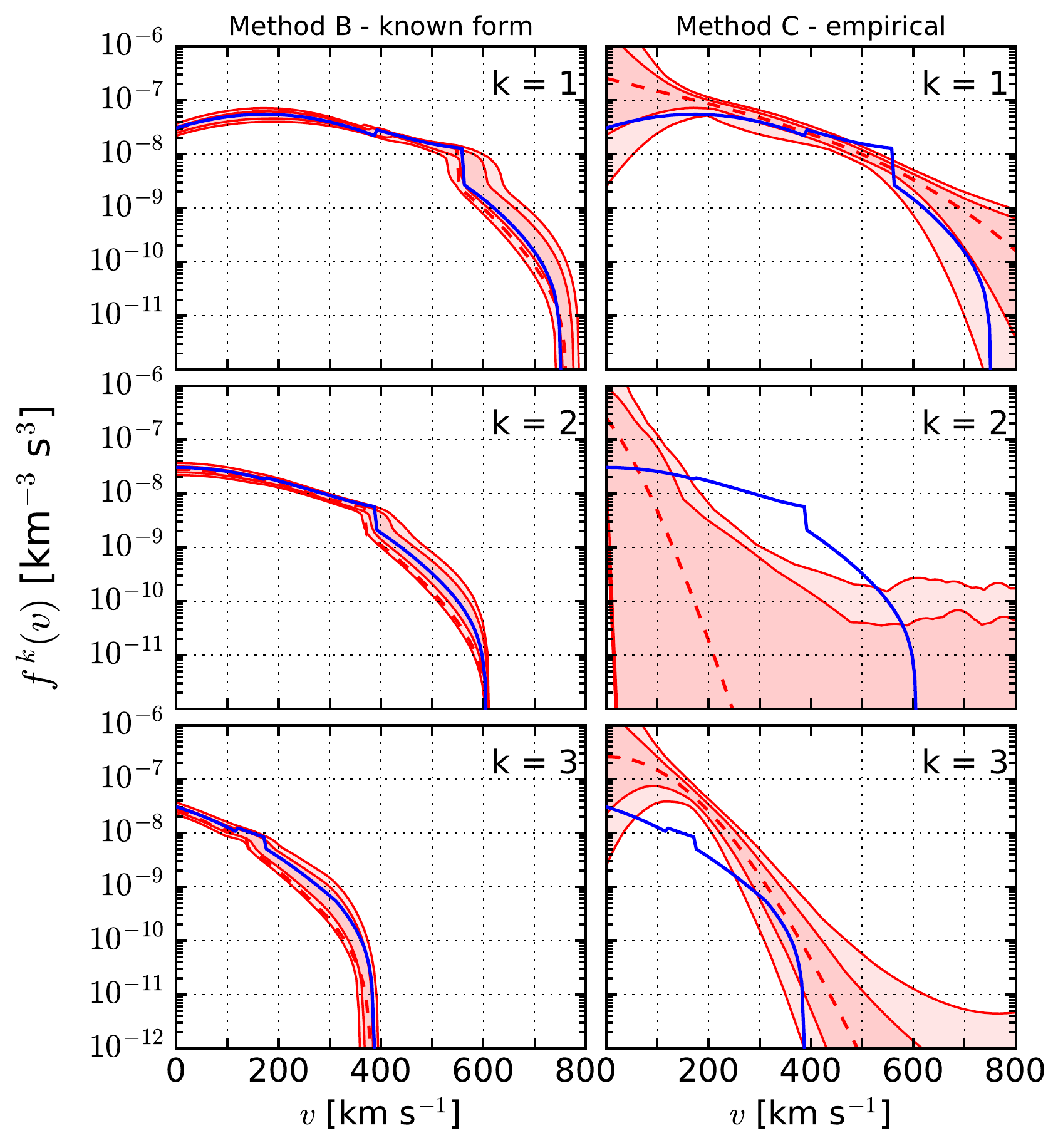}}
    \caption{Reconstructed velocity distribution averaged over each of the three angular bins ($k=1,2,3$) defined in Eq.~\ref{eq:bins}. The left column in each figure shows the results for Method B (known functional form) while the right column shows results for Method C (empirical parametrisation). The correct underlying velocity distribution is specified in the caption and shown as a solid blue line. The best fit reconstruction is shown as a red dashed line, while the 68\% and 95\% intervals are given by the inner and outer red shaded regions. The top left figure (a) shows results in which only the Fluorine-based experiment has directional sensitivity. In the remaining figures, both Fluorine and Xenon experiments are directionally sensitive.}\label{fig:veldist}
\end{figure*}

We now present results for the shape of the reconstructed velocity distribution in each of the three angular bins:
\begin{align}\label{eq:bins}
\begin{split}
k = 1:& \qquad \theta \in \left[ 0, \pi/3\right]\,, \\
k = 2:& \qquad \theta \in \left[ \pi/3, 2\pi/3 \right]\,, \\
k = 3:& \qquad \theta \in \left[ 2\pi/3, \pi\right]\,. \\
\end{split}
\end{align}
For the discretised velocity distribution of Method C, we simply construct the velocity distribution in the $k$th bin, $f^k(v)$, from the $\{a_m^{(k)}\}$ parameters according to Eq.~\ref{eq:polynomialparam}. For Method B, we average the full velocity distribution (described by a given set of parameters) over each angular bin in $k$:
\begin{equation}\label{eq:fk}
f^{k}(v) = \frac{\int_{\cos(k\pi/N)}^{\cos((k-1)\pi/N)} f(\mathbf{v}) \, \mathrm{d}\cos\theta}{\cos((k-1)\pi/N) - \cos(k\pi/N)}\,.
\end{equation}
At each speed $v$, 68\% and 95\% confidence intervals are calculated from the distribution of values of $f^k(v)$ by profiling over the values at all other speeds (and profiling over the mass and cross section). Figure~\ref{fig:veldist} compares the reconstructed distributions $f^k(v)$ in the two Methods B and C (red curves) as well as `true' distributions obtained by applying Eq.~\ref{eq:fk} to the correct underlying distribution (solid blue curve).

Figure~\ref{fig:veldist-a} shows results for the SHM distribution with directional sensitivity in only the Fluorine experiment. For Method B (left column), the best fit velocity distribution (dashed red) follows closely the underlying distribution, with narrow confidence intervals (shaded red bands). The strongest constraints are in the forward bin ($k=1$) in the range $v \sim 300 \textendash 500 \kms$. This is due to the distribution of recoils which is focused in the forward direction, with the rate of recoils peaking a little above the energy threshold of the Fluorine detector (corresponding to a speed of $v \sim 300 \kms$ for a DM mass of 50 GeV). 

Using the empirical parametrisation (right column), we also obtain a good fit to the velocity distribution in the forward bin. At high and low speeds, the confidence intervals widen as the recoil rate is insensitive to the shape of the shape of the speed distribution outside of the energy window $[E_\mathrm{th}, E_\mathrm{max}]$ of the analysis. In the transverse ($k=2$) and backward ($k=3$) bins, the velocity distribution is also poorly constrained, with no lower limit over the full range of speeds. The $k=3$ velocity distribution contributes predominantly to recoils in the backwards direction. There are zero backward-going events in the Fluorine dataset, leading to poor constraints. 

Comparing now with Fig.~\ref{fig:veldist-b}, in which both Xe and F detectors have directional sensitivity, we see that the constraints are tightened. For Method B, this is perhaps most pronounced for the $k = 3$ bin. The lower threshold of the Xenon detector (compared to the Fluorine one) produces a distribution of nuclear recoils which is less strongly peaked in the forward direction\footnote{The SHM distribution (in the Earth frame) is increasingly anisotropic with increasing speed $v$. A lower energy threshold leads to a smaller value of $v_\mathrm{min}$ and therefore allows the experiment to access the lower speed, more isotropic part of the velocity distribution.}, meaning that more recoils are observed in the backward and transverse direction, improving constraints in all three velocity bins.  

Similarly, constraints in the $k=3$ bin for Method C are also now stronger, with closed confidence intervals at both the 68\% and 95\% levels. However, the best fit velocity distribution in this bin appears to be slightly larger than the true distribution. In contrast, the discretised velocity distribution in the $k=2$ bin is significantly lower than the true distribution averaged over that bin. As is clear from the top panel of Fig.~\ref{fig:polar}, $f(\mathbf{v})$ is in fact a strong function of $\theta$ across the $k=2$ bin. If we fixed $f^2(v)$ equal to the average of the true distribution across the entire bin, this would lead to an excess of recoils in the backwards direction and a deficit of recoils in the forward direction. Instead, the best fit form of $f^2(v)$ peaks at low speeds, with only a small contribution above the experimental thresholds. There is then still sufficient freedom in $f^1(v)$ and $f^3(v)$ to fit the observed distribution of recoils. 

We now consider the reconstructions of $f^k(v)$ for the two alternative halo models. Fig.~\ref{fig:veldist-c} shows results for the SHM+Str model when both experiments are directionally sensitive. As before, when the underlying functional form is known (left column), the velocity distribution is well reconstructed, with the stream being tightly constrained in this case.  For the empirical parametrisation (right column), the 4-parameter polynomial fit in each angular bin is not sufficient to pick out a feature as sharp as a stream. Nonetheless, the reconstruction does point towards an excess of particles in the $k=2$ bin in a wide range around the stream speed of $400 \kms$. In contrast to the SHM-only benchmark, the stream leads to an enhanced rate in the transverse recoil direction, which requires a significant $k=2$ population to match the observed recoil distribution. To compensate, the best fit form of $f^3(k)$ is suppressed, though in all 3 bins the underlying distribution falls within the 95\% intervals.

Finally, we consider the SHM+DF model in Fig.~\ref{fig:veldist-d}. For Method B, the confidence intervals are slightly wider than in the case of the SHM+Str. This is because the Debris Flow is a broader feature in the velocity distribution (see the bottom panel of Fig.~\ref{fig:polar}) and therefore has a stronger degeneracy with the parameters of the SHM. For Method C, we see a slightly flatter reconstructed distribution in the $k=1$ bin than for previous benchmarks, as well as narrower uncertainty bands up to around $550 \kms$. This is a result of the enhancement in high energy recoils in the forward direction, caused by the high speed debris flow. 

In this section, we have observed that the discretised, empirical parametrisation of Method C can provide a close approximation to the shape of the (bin-averaged) velocity distribution in some scenarios, in particular when large numbers of events are observed in a particular direction. In other cases, there appears to be a discrepancy between the reconstructions and the underlying distribution. However, this is only a problem if we interpret the $f^k(v)$ of Eq.~\ref{eq:polynomialparam} as representing the average of the true speed distribution across the $k$th bin (as defined in Eq.~\ref{eq:fk} and illustrated by the blue curves). As discussed above, setting $f^k(v)$ equal to the bin-averaged velocity distribution does not necessarily provide a good approximation to the full velocity distribution. Instead, we should interpret the $f^k(v)$ functions as empirical fits to the full velocity distribution. These can be used to look for clear features in the DM distribution (for example, the stream population in Fig.~\ref{fig:veldist-c}), but it is difficult to make statistically concrete statements about the underlying velocity distribution from the shapes of $f^k(v)$. In the next section, we discuss some simple measures which can be used to extract information and compare different possible velocity distributions.

\subsection{Velocity parameters}
\label{sec:parameters}

\begin{figure*}
	\includegraphics[width=0.32\textwidth]{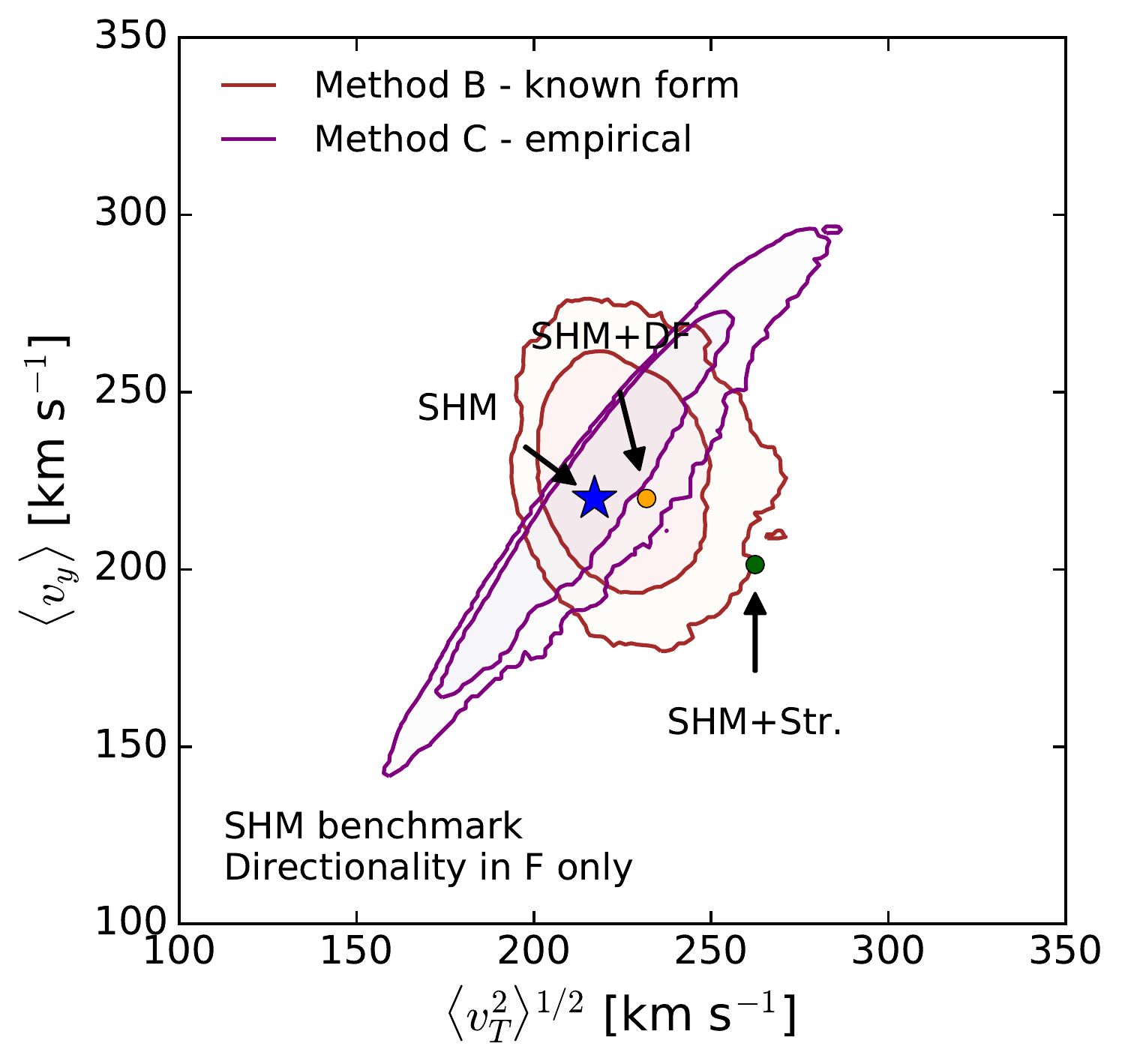}
	\includegraphics[width=0.32\textwidth]{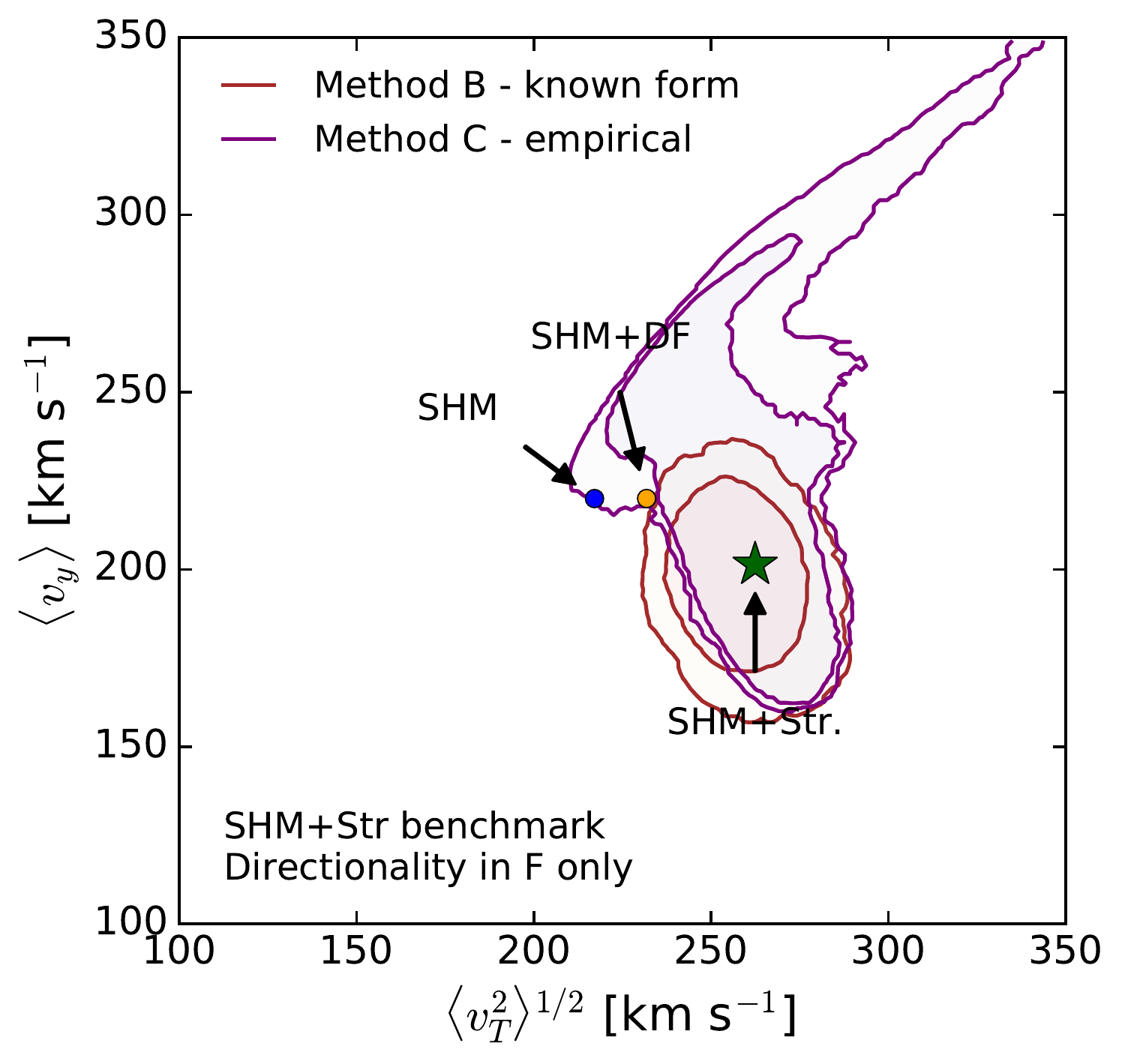}
	\includegraphics[width=0.32\textwidth]{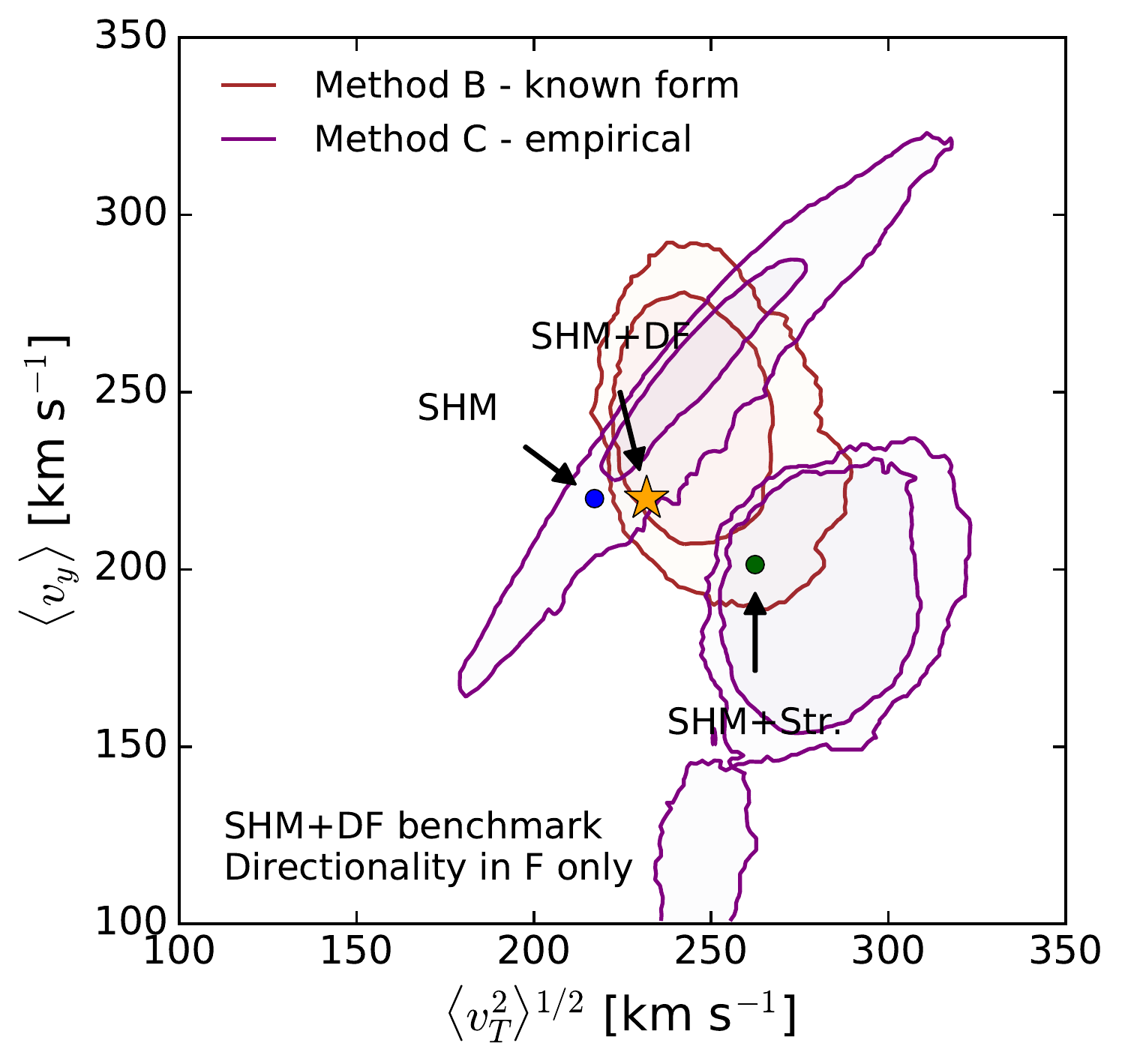}
	\includegraphics[width=0.32\textwidth]{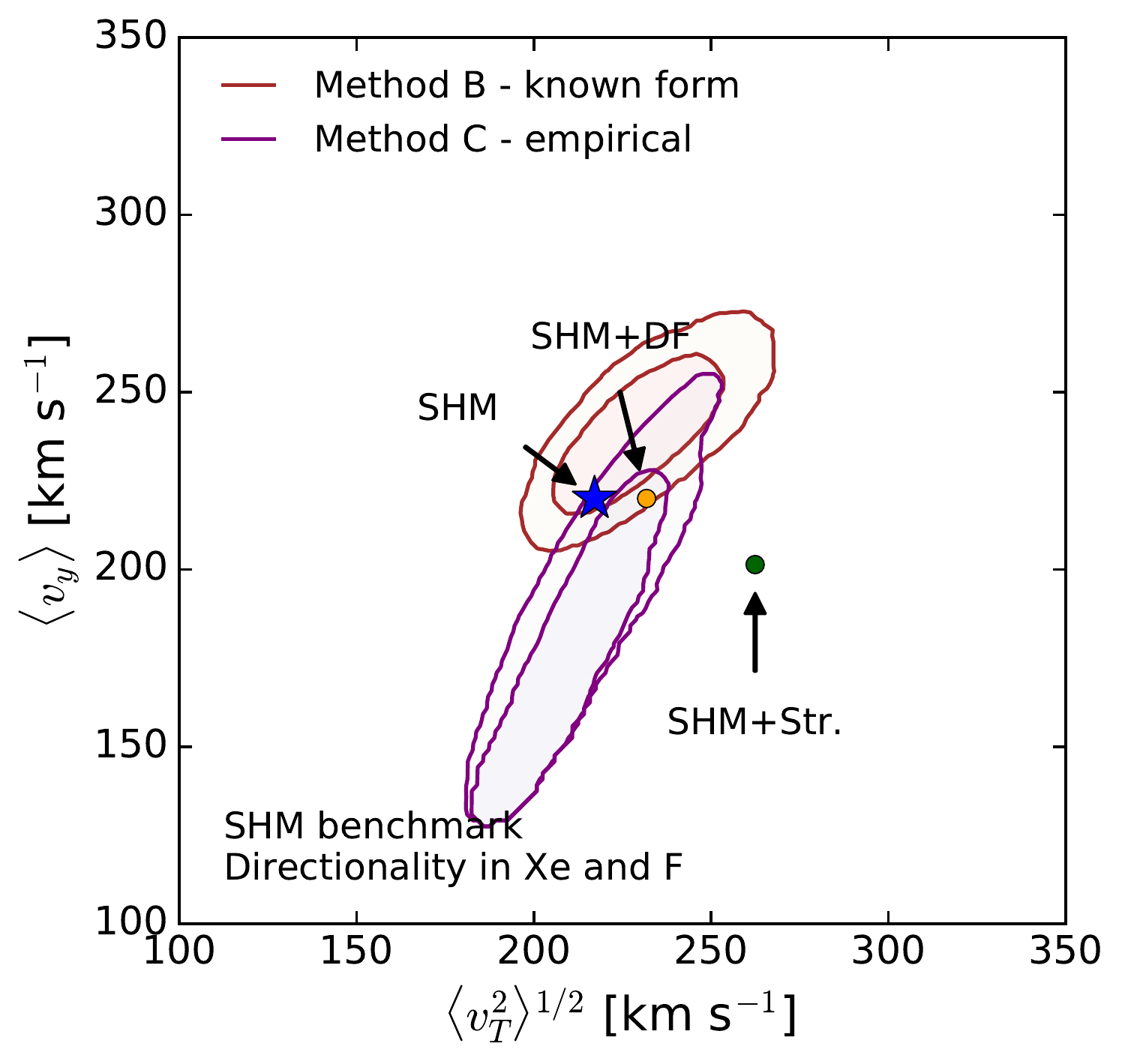}
	\includegraphics[width=0.32\textwidth]{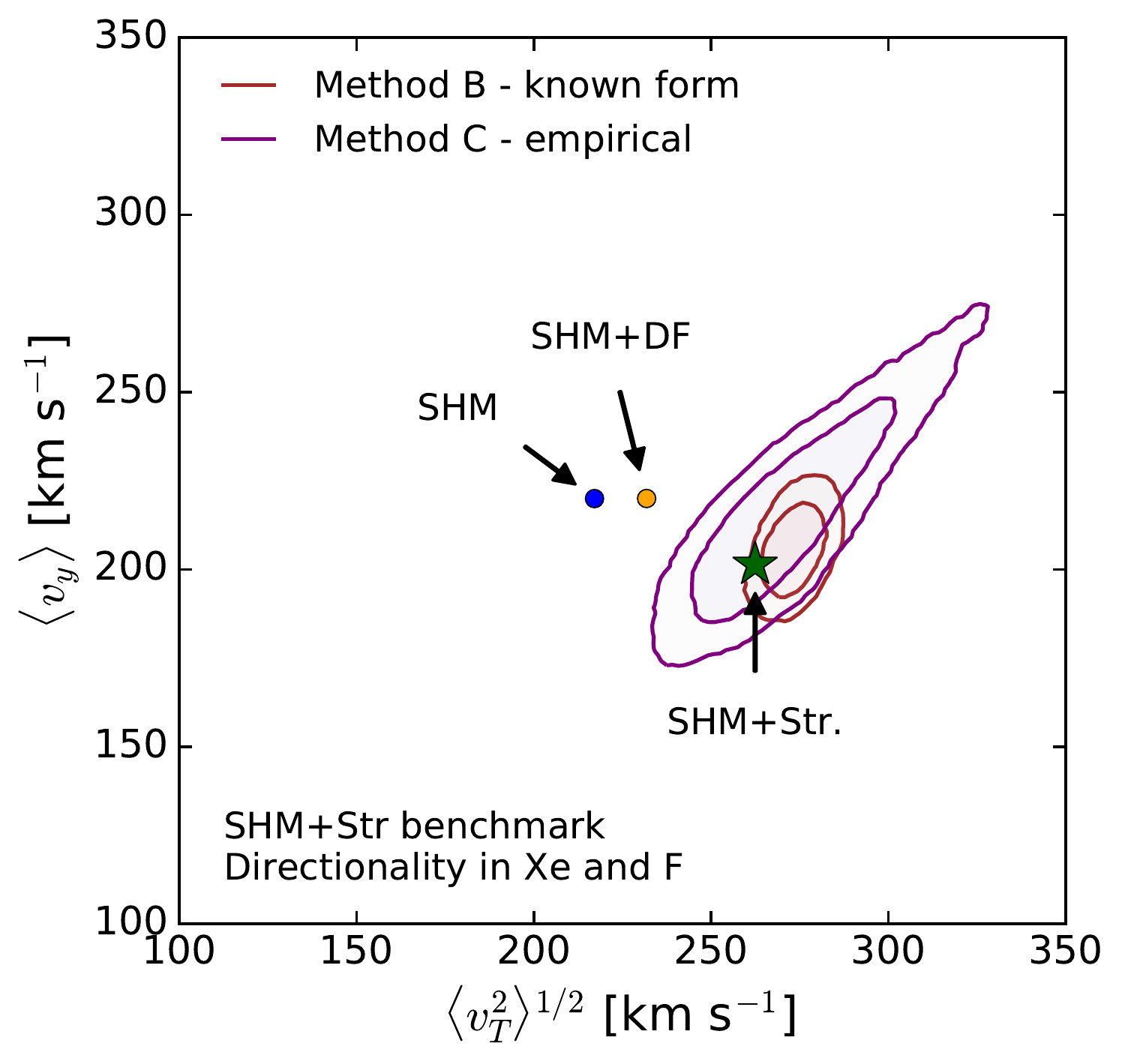}
	\includegraphics[width=0.32\textwidth]{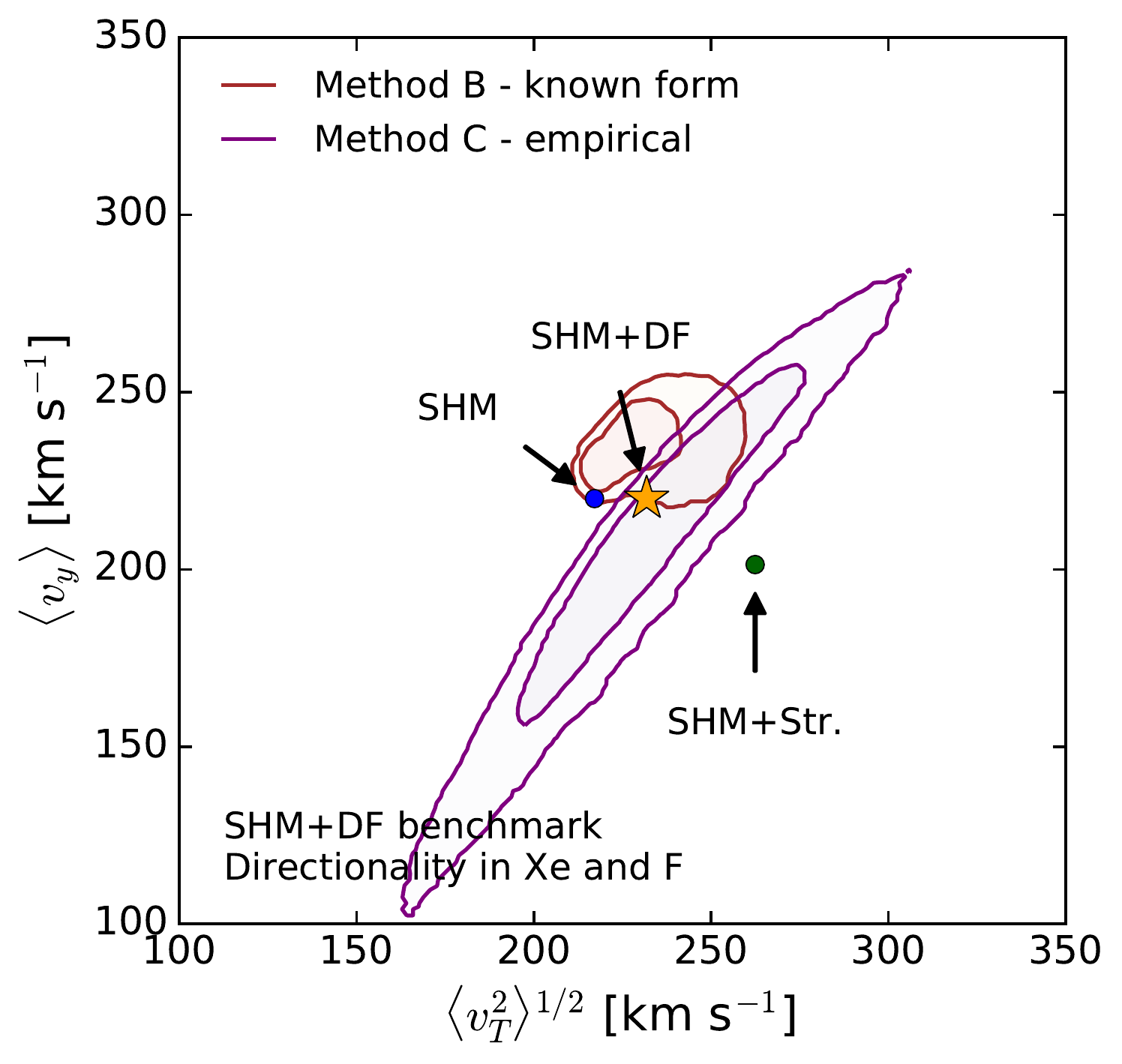}
    \caption{Mean values for the DM velocity parallel to the Earth's direction $\langle v_y \rangle$ and the transverse velocity perpendicular to the Earth's motion, $\langle v_T^2 \rangle^{1/2}$. The 68\% and 95\% confidence intervals obtained using reconstruction methods B and C are shown as pairs of red and purple contours respectively. We show results for all three halo models (SHM, SHM+Str and SHM+DF in each column from left to right) and for directionality in a single experiment (top row) and in both experiments (bottom). For each benchmark, the correct value of $\langle v_y \rangle$ and $\langle v_T^2 \rangle^{1/2}$ is marked by a large star, while the values for alternative halo models are shown as small circles.}\label{fig:vyvT}
\end{figure*}

Given that direct detection experiments are the only way to probe the DM velocity distribution down to sub-milliparsec scales, a central goal of the post-discovery era will be to determine the quantity of substructure in the local DM halo. Because substructures can give rise to phenomenologically varied signatures in recoil spectra it will be useful to attempt to discriminate between different classes of substructure in a model-independent way. Attempts have previously been made to use non-parametric statistics in directional experiments to search for substructure or anisotropies in the velocity distribution~\cite{Morgan:2004ys,Green:2010zm,O'Hare:2014oxa}. However these tests do not allow all of the properties of the substructure to be measured and require much larger numbers of events to be successful. 

We can discriminate between our three halo models in a simple way by mapping the reconstructions as presented in the previous section on to a set of physical parameters that can be extracted by both methods (B and C) for fitting the velocity distribution. We calculate mean values for the velocity parallel and transverse to the Earth's motion, $\langle v_y\rangle$ and $\langle v_T^2 \rangle$ respectively:
\begin{equation}\label{eq:vy}
\langle v_y \rangle = \int \mathrm{d}v \,\int_{0}^{2\pi} \mathrm{d}\phi \, \int_{-1}^1 \mathrm{d}\cos\theta \, (v\cos\theta)\, v^2 f(\mathbf{v}) \, ,
\end{equation}
and
\begin{equation}\label{eq:vT}
\langle v_T^2 \rangle = \int \mathrm{d}v \,\int_{0}^{2\pi} \mathrm{d}\phi \, \int_{-1}^1 \mathrm{d}\cos\theta \, (v^2(1-\cos^2\theta))\, v^2 f(\mathbf{v}) \, .
\end{equation}

In Fig.~\ref{fig:vyvT} we show the reconstructed velocity distribution in each halo model mapped on to the $\langle v_y \rangle$-$\sqrt{\langle v_T^2 \rangle}$ plane. Here again we make the comparison between only one experiment having directional sensitivity (F) and both experiments being directionally sensitive. For Method B (red), the values of the physical parameters ($v_0$, $\sigma_v$, $v_s$, etc.) are typically well constrained, meaning that $\vy$ and $\vT$ are also well constrained, with roughly Gaussian error contours \footnote{In fact, in some cases the fit parameters and the derived parameters are closely related. For example, in the SHM, there is a close correspondence between $v_0$ and $\vy$.}.  In contrast, the reconstructions using Method C (purple) exhibit a pronounced degeneracy along the direction of $\vy \propto \vT$ for many of the benchmarks. This is due to the fact that the $k=1$ (forward) and $k=3$ (backward) bins contribute to the mean values of both the forward \textit{and} transverse DM speeds. For example, increasing $f^1(v)$ in the forward bin leads to an increase in $\vy$ but also a proportional increase in $\vT$, because the particles are assumed to be distributed equally in $\theta$ across the bin. The position of the contours in $\vT$ is typically dominated by the $k=2$ bin, which contributes only to $\vT$ and not to $\vy$. 

For both reconstruction methods and for directionality in either one or both experiments, the underlying benchmark values of $\vy$ and $\vT$ always lie within the 95\% confidence regions. The SHM and SHM+DF models are hardest to distinguish. The debris flow is isotropic in the Galactic frame, so the net velocity of the DM particles in the Lab frame is due entirely to the Earth's motion. Thus, we have $\vy \sim v_0$ as in the SHM. Furthermore, as can be seen in the middle panel of Fig.~\ref{fig:polar}, the debris flow is rather broad (rather than being focused in one particular direction), leading only to a mild increase in $\vT$. Indeed, with the SHM benchmark dataset, the SHM+DF model cannot be rejected at the 95\% confidence level using either reconstruction method.  

The SHM+Str is much more easily distinguished from the other two benchmarks. The stream velocity is almost perpendicular to the Earth's velocity, leading to a decrease in $\vy$ and a marked increase in $\vT$. For the SHM mock dataset, the SHM+Str is clearly excluded, even when only the Fluorine detector has directionality. Conversely, when using the SHM+Str mock dataset (middle column), the SHM and SHM+DF are excluded at the 95\% level when the true functional form is known (Method B), but lie close to the 95\% contour when the empirical form is used (Method C) and only Fluorine has directional sensitivity. The addition of the Stream component leads to a mild increase in the number of Fluorine events in the transverse recoil direction (relative to the SHM alone), while still producing no events in the backward recoil direction. This data can be well fit by adding a substantial population of particles in the $k=2$ bin (the lower, round part of the contours in the upper middle panel of Fig.~\ref{fig:vyvT}) or by enhancing the forward $k=1$ population, particularly at low speeds, which are more likely to produce transverse recoils (the upper, straight part of the contours). In Xenon, the numbers of forward and transverse recoils are roughly equal, which breaks this degeneracy (lower middle panel of Fig.~\ref{fig:vyvT}) and allows the SHM+Str model to be unequivocally distinguished from the other benchmarks.

Using the SHM+DF dataset  with directionality in Fluorine only (upper right panel of Fig.~\ref{fig:vyvT}), there are now three distinct regions which fit the data using Method C. The three regions correspond to enhanced populations of DM particles in the $k=1$, $k=2$ and $k=3$ bins (from top to bottom respectively). It is clear from Fig.~\ref{fig:polar} that the Debris Flow contributes in all 3 angular bins and will typically produce higher energy recoils than the SHM alone (as $v_f > v_0$). An increased high-speed population in any of the three velocity bins (relative to the smooth SHM) will then improve the fit to the data. Once again, adding the Xenon detector (with its different directional spectrum) breaks the degeneracy between the three regions and in this case the SHM+Str benchmark can be rejected in both Methods B and C (lower right panel of Fig.~\ref{fig:vyvT}) . 

These results indicate that mapping the reconstructed velocity distributions onto the parameters $\vy$ and $\vT$ can be a reliable and unbiased way of trying to distinguish different underlying halo models. The SHM and SHM+DF models are typically difficult to distinguish, while the SHM+Str has sufficiently different properties (in particular a large transverse velocity component) that it can be clearly excluded in many cases. The Xenon detector we have considered has a slightly more isotropic distribution of recoils than the Fluorine detector (due to its low threshold). This allows us to break certain degeneracies between the different angular bins, as well as to tighten the overall constraints. Indeed, with directionality in both detectors, the correct benchmark values of  $\vy$ and $\vT$ are recovered in all three models.

\subsection{Folded reconstructions}
\label{sec:folded}
\begin{figure*}[t]
	\includegraphics[width=0.32\textwidth]{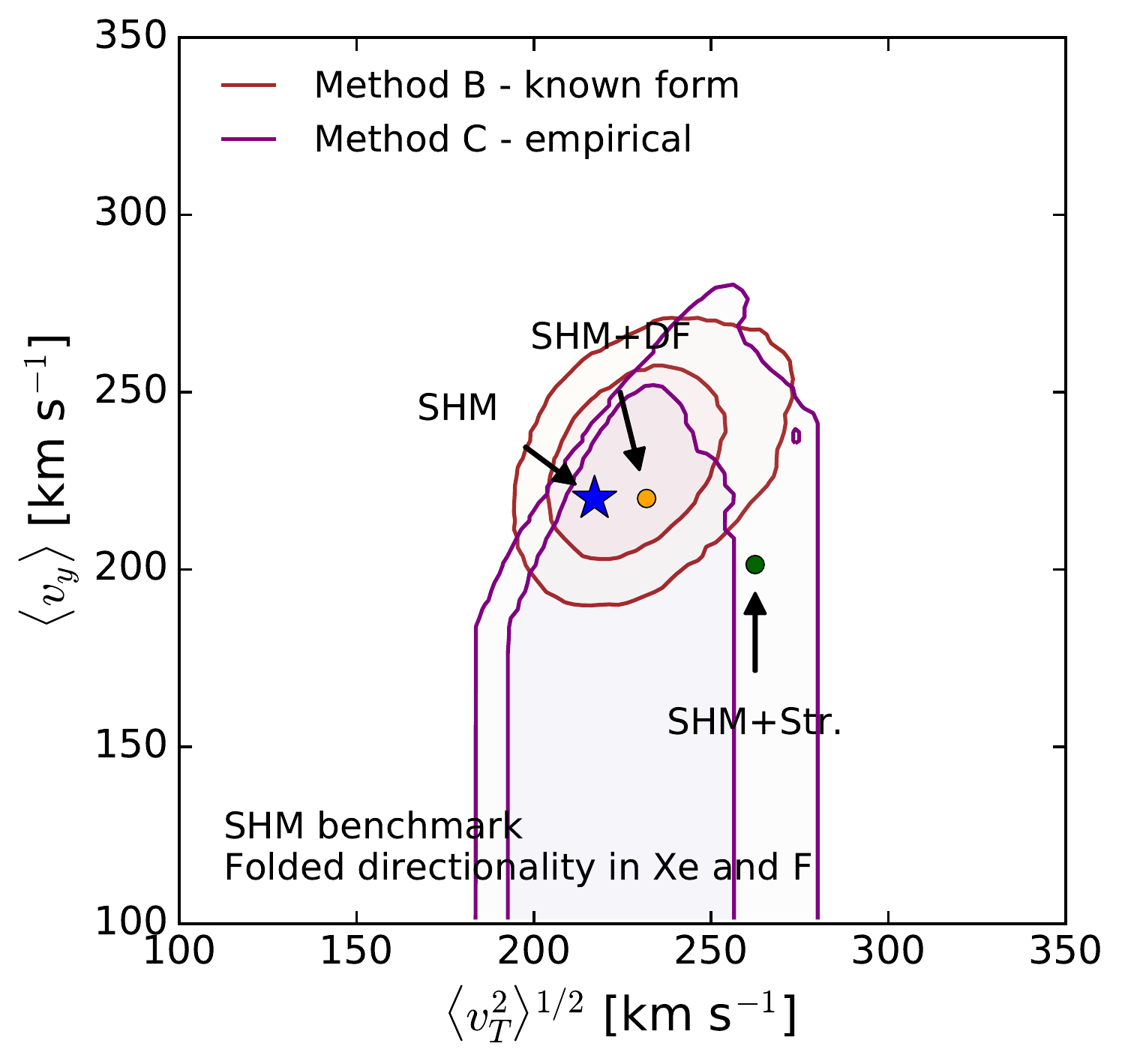}
	\includegraphics[width=0.32\textwidth]{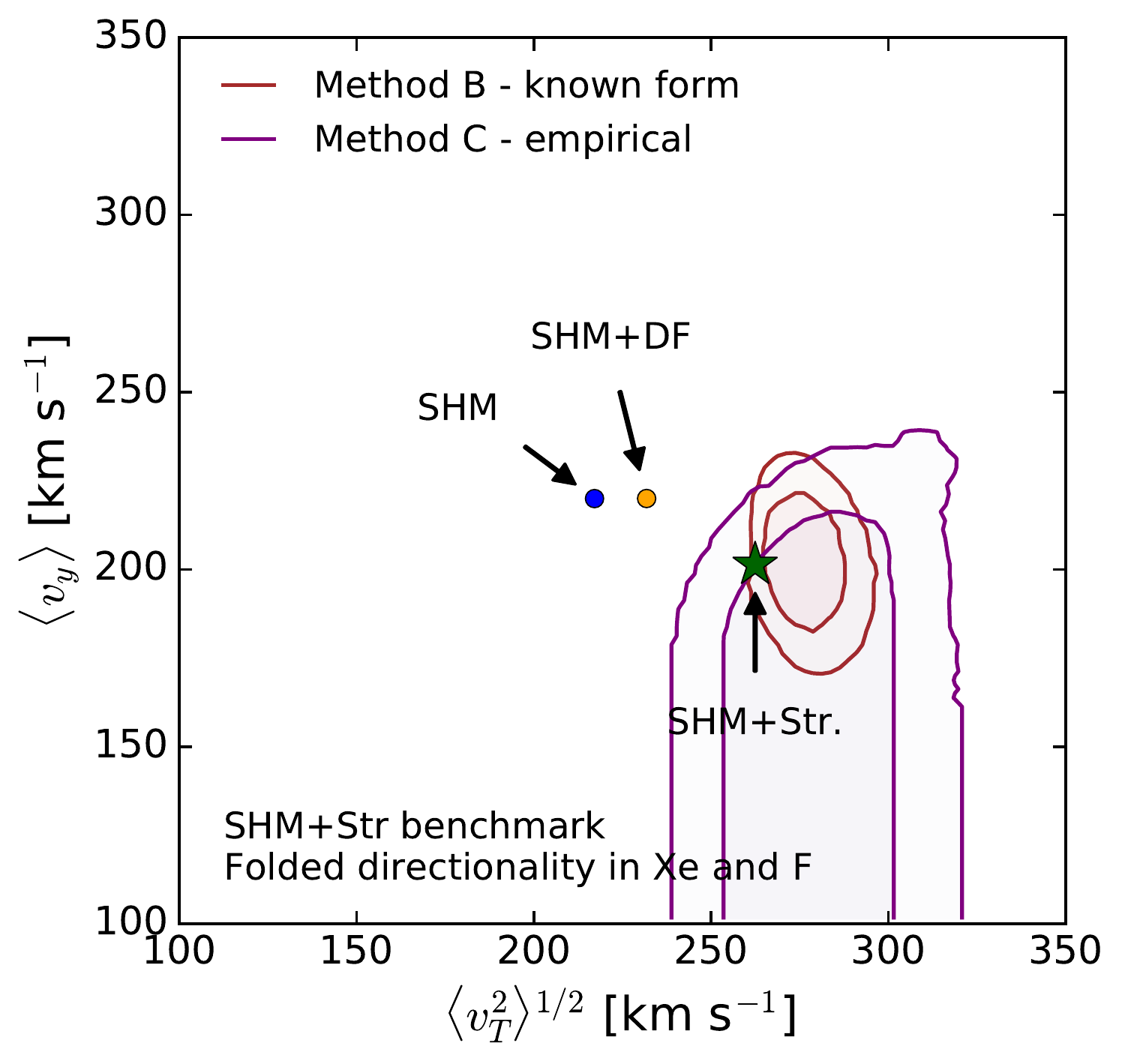}
	\includegraphics[width=0.32\textwidth]{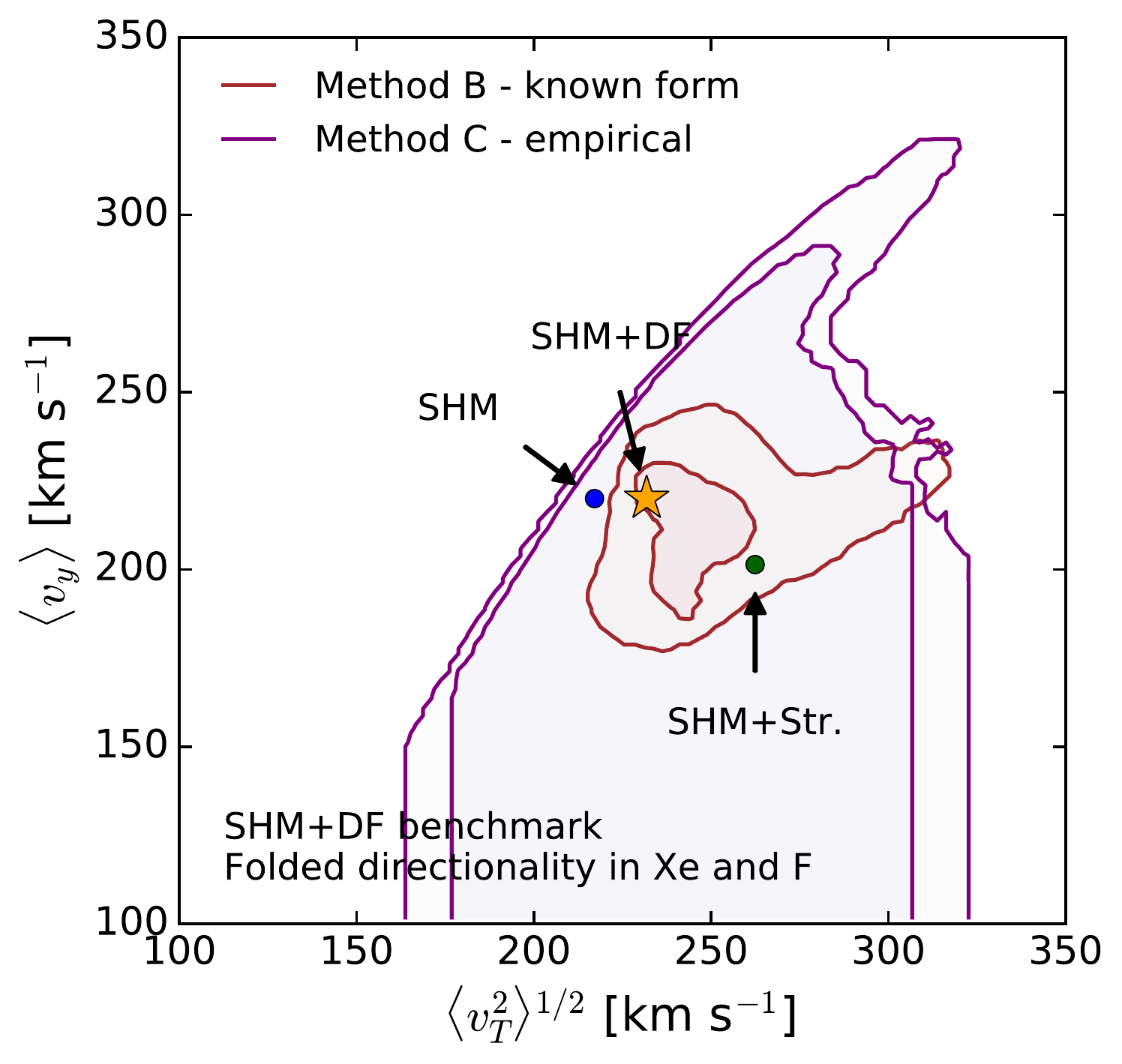}
    \caption{Mean values for the DM velocity parallel to the Earth's direction $\langle v_y \rangle$ and the transverse velocity perpendicular to the Earth's motion, $\langle v_T^2 \rangle^{1/2}$, reconstructed when both experiments have directional sensitivity but lack any sense recognition. The 68\% and 95\% confidence intervals obtained using reconstruction methods B and C are shown as pairs of red and purple contours respectively. We show the results for each halo model (from left to right), the SHM, the SHM+Str and SHM+Debris flow models. For Method C (purple), the contours extend all the way down to negative values of $\vy$, but for clarity we show only the region of parameter space near the benchmark values. For each benchmark, the correct value of $\langle v_y \rangle$ and $\langle v_T^2 \rangle^{1/2}$ is marked by a large star, while the values for alternative halo models are shown as small circles.}\label{fig:vyvT-folded}
\end{figure*}

A major concern for current directional detection experiments is the ability to measure the forward or backward going sense of a reconstructed recoil track. In standard low pressure gas TPC experiments head-tail recognition is achievable if there is a measurement of any asymmetry in either the angular dispersion or charge deposition along the track~\cite{Billard:2012bk}. Head-tail asymmetry in nuclear recoils has been observed experimentally~\cite{Majewski:2009an,Dujmic:2007bd,Burgos:2008jm,Battat:2016xaw}. However given that the lack sense recognition is a significant limitation of directional detectors at DM recoil energies, we now present reconstructions of the forward-backward folding of the velocity distribution.

We define the `folded' recoil spectrum that would be observed in experiments without any head-tail effect as,
\begin{equation}
 \frac{\mathrm{d}^2R_\mathrm{fold}}{\mathrm{d}E_r\mathrm{d}\Omega_q} = \frac{\mathrm{d}^2R}{\mathrm{d}E_r\mathrm{d}\Omega_q}\bigg|_{-\qhat}+\frac{\mathrm{d}^2R}{\mathrm{d}E_r\mathrm{d}\Omega_q}\bigg|_{+\qhat} \, .
\end{equation}

Following the results of Sec.~\ref{sec:parameters} we show again the expectation values for the parallel and transverse velocities with respect to the direction of the Earth's motion. In this case for brevity we include only the result for the case in which both F and Xe experiments have directional sensitivity, only now we remove their ability to tell the forward or backward going sense of their nuclear recoils. The results are shown for each halo model in Fig.~\ref{fig:vyvT-folded}.

With the removal of sense recognition the pronounced dipole feature of the angular distribution of recoils is reduced. Hence our directional experiments can no longer extract information about the asymmetry between forward and backward going recoils. For Method B there is only a small increase in the size of the contours for the SHM and SHM+Str models as in these cases there are large populations of recoils transverse to the folding so there is not a large reduction in sensitivity to the parameters that are being reconstructed. Whilst there is a larger uncertainty in the full 3-dimensional stream velocity in Method B, this uncertainty is disguised by the mapping onto $\vy$ and $\vT$ and the SHM and SHM+Str benchmarks can still be distinguished. However in the case of the SHM+DF model there is a moderate increase in the size of the contours in $\vy-\vT$. This is because some of the information regarding the velocity of the debris flow is encoded in the forward-backward asymmetry of the recoils.

For Method C, however, we see a complete degeneracy appearing in the results for all three halo models between positive and negative values of $v_y$ (although for clarity we display only positive values of $v_y$ here). This is to be expected as the folded distribution measured by Method C has no distinction between $\vy$ running parallel or anti-parallel to the Earth's motion. However as we have not removed any transverse velocity information, the shape of the contours in the $\vT$ direction remain relatively unchanged for the SHM and SHM+Str models. In particular, for data under the SHM+Str benchmark, the SHM and SHM+DF benchmarks can still be rejected at the 95\% confidence level. However, this is not the case for the SHM+DF model; the debris flow component has populations in both transverse and parallel directions so there is a significant increase in the size of the countours in both $\vy$ and $\vT$. In this case all three benchmarks lie within the 68\% region.

\section{Conclusions}
\label{sec:conc}

We have explored a number of methods for reconstructing the DM velocity distribution from future directional experiments. We have focused in particular on using a general, empirical parametrisation to fit the velocity distribution and compared this with the case where the underlying form of the velocity distribution is known. This allows us to understand whether the two methods lead to different reconstructed parameter values (which may be indicative of biased reconstructions) and how much the constraining power of the experiments changes as we open up the parameter space with a more general fit.

Previous works have demonstrated that the DM mass can be recovered from non-directional direct detection experiments without making assumptions about the form of the speed distribution \cite{Peter:2011eu,Kavanagh:2013wba}. As we show in Fig.~\ref{fig:mx-recon}, such astrophysics-independent approaches can be successfully extended to directional experiments. In particular, the use of an approximate, discretised velocity distribution does not spoil the accurate reconstruction of the DM mass. Our empirical parametrisation typically leads to larger uncertainties than when the underlying form of the distribution is known, but we see no evidence of bias. The DM mass reconstructed using the two methods is similar in almost all cases and the true DM mass of 50 GeV is always enclosed within the $95\%$ confidence intervals over a range of halo models.

In principle, we should also be able to recover the DM velocity distribution as well as the DM mass. In order to make the fitting procedure tractable, we have discretised the velocity distribution into $N=3$ distinct angular bins. As demonstrated in Sec.~\ref{sec:shape}, looking at the speed distribution $f^k(v)$ within each angular bin may allow us to pick out key features but it is generally difficult to make comparisons with different possible underlying velocity distributions. Instead, we construct confidence intervals for $\vy$ and $\vT$, the average DM velocity parallel and transverse to the direction of the Earth's motion. These measures of the shape of the distribution allow us to distinguish robustly between different underlying halo models. Although a perfect reconstruction of the full velocity distribution is difficult even with large event numbers, we have shown that this model independent approach can be used as a first step in identifying deviations from the assumption of the SHM to point towards the existence of substructures. In principle one could then move to a particular model dependent parametrisation which would be able to measure the substructure more accurately and extract the astrophysically meaningful parameters.

We find that with directionality only in a Fluorine experiment, it may be possible to detect or reject the presence of a substantial stream with 95\% confidence. More isotropic features, such as a debris flow, are more difficult to distinguish from the SHM. Adding directionality in a Xenon experiment allows us to break degeneracies in the shape of the velocity distribution and leads to good discrimination between models with and without a stream. The SHM and SHM+DF models remain harder to distinguish using this method, whether the underlying functional form is known or not.

In experiments without the ability to determine the sense of the nuclear recoils we see the discretised approach suffer. This is because the $N=3$ binning is effectively reduced to 2 as the forward and backward bins are folded. The result of this is that it becomes impossible to precisely measure the average speed in the direction of the folding due to a degeneracy between positive and negative values. This confirms the results of previous studies~\cite{Green:2007at,Billard:2014ewa} finding that the lack of sense recognition greatly reduces the power of directional experiments.

The benchmark examples we have chosen in this work enable us to broadly compare the success of a discretised parametrisation of the DM velocity distribution under a range of scenarios. However the parameter space that describes different classes of substructure, for instance streams, is large. It is unlikely that the conclusions drawn from our benchmark (which includes a rather large stream component) can be extended generally over the range of possible stream speeds and directions. However, we have demonstrated that an empirical parametrisation can accommodate a wide range of underlying velocity distributions without a large loss in sensitivity compared to when the functional form is fixed and known.

In this work, we have considered only ideal direct detection experiments. Experimental complications such as finite energy and angular resolution, as well as the possibility of lower-dimensional readouts, will of course affect the reconstruction of the DM parameters in real experiments. We note, however, that the angular binning procedure we have used in the empirical reconstructions may be a natural way to account for finite angular resolution. If the angular resolution (typically in the range 20$\,^{\circ}$-80$\,^{\circ}$ \cite{Billard:2012bk}) is smaller than the binning angle (here, $60^\circ$), the inclusion of these effects should have little impact on the results. It is not yet clear, however, what the optimum binning angle (and therefore the optimum number of bins) would be.

In spite of these open questions, the study we have presented here shows that for exploring the full three-dimensional local velocity distribution, which is a primary motivation for directional experiments, one can make significant progress without assumptions about the underlying astrophysics. The method we have presented allows one to combine directional and non-directional experiments in a general way in order to accurately reconstruct the DM mass, identify broad features in the DM velocity distribution and perhaps even distinguish different underlying models for the DM halo.

\acknowledgments

The authors thank Anne Green for helpful comments on this manuscript. BJK is supported by the John Templeton Foundation Grant 48222 and by the European Research Council ({\sc Erc}) under the EU Seventh Framework Programme (FP7/2007-2013)/{\sc Erc} Starting Grant (agreement n.\ 278234 --- `{\sc NewDark}' project). BJK also acknowledges the hospitality of the Institut d'Astrophysique de Paris, where part of this work was completed. CAJO is supported by a UK Science and Technology Facilities Council studentship. Access to the High Performance Computing facilities provided by the Institut des Sciences du Calcul et des Donn\'ees (Sorbonne Universit\'es) is gratefully acknowledged. 
\bibliography{DirectionalDetailed.bib}

\end{document}